\documentstyle[11pt,epsfig]{article}

\newlength{\bredde}
\def\slash#1{\settowidth{\bredde}{$#1$}\ifmmode\,\raisebox{.15ex}{/}
\hspace*{-\bredde} #1\else$\,\raisebox{.15ex}{/}\hspace*{-\bredde} #1$\fi}
\newcommand{\beq}{\begin{equation}}
\newcommand{\eeq}{\end{equation}}
\newcommand{\ba}{\begin{array}{ccc}}
\newcommand{\ea}{\end{array}}

\newcommand{\noi}{\vspace{12pt}\noindent}
\newcommand{\lG}{\raise.3ex\hbox{$\stackrel{\leftarrow}{G}$}}
\newcommand{\lU}{\raise.3ex\hbox{$\stackrel{\leftarrow}{U}$}}
\newcommand{\lP}{\raise.3ex\hbox{$\stackrel{\leftarrow}{{\cal P}}$}}
\newcommand{\leta}{\raise.3ex\hbox{$\stackrel{\leftarrow}{\eta}$}}
\newcommand{\lOmega}{\raise.3ex\hbox{$\stackrel{\leftarrow}{\Omega}$}}
\newcommand{\ldr}{\raise.3ex\hbox{$\stackrel{\leftarrow}{\delta^r}$}}

\def\m2{{\mathcal{M}}^{\dagger}{\mathcal{M}}}
\def\mb2{M^2}

\def\beqn{\begin{eqnarray}}
\def\eeqn{\end{eqnarray}}

\def\gtwid{\raise.3ex\hbox{$>$\kern-.75em\lower1ex\hbox{$\sim$}}}
\def\ltwid{\raise.3ex\hbox{$<$\kern-.75em\lower1ex\hbox{$\sim$}}}

\def\Tr{ {\rm Tr} }

\begin{document}
\topmargin -1.4cm
\oddsidemargin -0.8cm
\evensidemargin -0.8cm

\begin{flushright}
{CERN-TH/2001-351}\\
{DESY-01-205} \\
{IFIC/01-66} \\
{FTUV-01-1212}\\
\end{flushright}
\vspace*{1cm}
\begin{center}
{\Large{\bf Finite-size Scaling of Meson Propagators}}\\

\vspace{1.cm}
P.H.~Damgaard$^{\rm a}$\footnote{ On leave
from Niels Bohr Institute, Blegdamsvej 17, DK-2100 Copenhagen, Denmark.}, M.C.~Diamantini$^{\rm a}$\footnote{Swiss National Science Foundation fellow.}, P.~Hern\'andez$^{\rm a}$\footnote{On leave
from Dept. de F\'{\i}sica Te\'orica, Universidad de Valencia.},
K.~Jansen$^{\rm b,}$

\vspace*{1cm}
$^{\rm a}$ Theory Division, CERN, 1211 Geneva 23, Switzerland \\
$^{\rm b}$ NIC/DESY Zeuthen, Platanenallee 6, D-15738 Zeuthen, Germany
\end{center}


\date{\today}
\vfill
\begin{abstract}
Using quenched chiral perturbation theory we compute meson correlation
functions at finite volume and fixed gauge field topology.
We also present the corresponding analytical predictions
for the unquenched theory at fixed gauge field topology.
These results can be used to measure the low-energy parameters of the 
chiral Langrangian from lattice simulations
in volumes much smaller than one pion Compton wavelength.
\end{abstract}
\vfill

\thispagestyle{empty}
\newpage

\setcounter{equation}{0}
\section{Introduction}

It is becoming increasingly clear that the study of strongly coupled gauge
theories in unphysical settings, such as finite volume or fixed gauge
field topology, can be extremely fruitful. From the point of view of lattice
simulations, the restriction to finite volume is of course ideal. It was
realized long ago that it can be advantageous to consider the propagation
of the pseudo-Goldstone bosons inside volumes too small to even
contain one Compton wavelength $1/m_{\pi}$
of these light degrees of freedom \cite{GL,N}.
This idea was explored in depth in the chiral Lagrangian framework of full
QCD \cite{H,HansenL}, as well as in a class of sigma models \cite{HL}.
Already at that time numerical simulations successfully reproduced these
predictions for sigma models \cite{HN,Hetal}.

\noi
In the meantime lattice simulations of QCD with very light quarks are
beginning to become feasible. If one restricts oneself to the same regime of
finite size $L \equiv V^{1/4}$, where $m_{\pi} \ll 1/L$, it was noted by
Leutwyler and Smilga that gauge field topology now plays a much more important
role than in the infinite-volume theory \cite{LS}. This fact can
be an advantage also in numerical simulations. If one has very definite
and different predictions in the various sectors of fixed topological
charge $\nu$, it means that one can perform a whole series of independent
fits to lattice data, and not just one.

\noi
It is thus of interest to combine finite volume with fixed gauge field
topology, and to consider QCD correlation functions in such a situation.
A systematic framework for analysing QCD in the regime $m_{\pi} \ll 1/L
\ll 4\pi F$,
the so-called $\epsilon$-expansion \cite{GL}, turns out to generalize
easily to sectors of fixed topology. In this paper we shall describe our
computation of
finite-volume meson correlation functions in the quenched analogue of this
$\epsilon$-expansion. We shall also provide new results for the unquenched
theory in the same finite-volume regime, but restricted to fixed topology.
The completely quenched theory is well-known to be special. Eliminating the
fermion determinant renders the theory non-unitary.
Still, it
is obviously of great interest to understand this truncation of the real
theory in view of lattice simulations. Here we shall adopt the
point of view that the quenched theory can be given some meaning close
to the full theory, over a certain range of scales. The difficulties of
the finite-volume quenched theory in the chiral limit at the quark level
show up again in the effective
field theory description, and there is no way around this fact.

\noi
Our paper is organized as follows. In section \ref{chiPT} we discuss the
two different ways in which to do quenched chiral perturbation theory,
the supersymmetric method, and the replica method. Section \ref{regimes}
discusses two different regimes of chiral perturbation
theory at finite volume, yielding two different perturbation series
known as the $p$- and $\epsilon$-expansions. We point out the peculiarities
of the counting of orders in these two expansions when going to the
quenched theory, and explain why sectors of fixed gauge field topology can be
particularly useful in the quenched case. In section \ref{chicondensate}
we recalculate the leading order correction to the quenched chiral
condensate in the $\epsilon$-expansion, and show that it gives the same
result as computed earlier in the replica formalism in \cite{D01}. Section \ref{QCF}
contains the results of our quenched calculation of mesonic correlation
functions at fixed topology, to leading order in the $\epsilon$-expansion.
We illustrate our results for typical values of lattice parameters, and
point out how these results can be used to determine the low-energy parameters of the chiral Lagrangian such as
the pion mass $m_{\pi}$ and the pion decay constant $F$. We also
present various checks on our calculations, such as consistency with
exact Ward identities. In section \ref{fullQCD} we present the results
of the analogous calculations for full QCD with $N_f$ light quark flavours
at fixed gauge field topology. Section \ref{conclude} contains our
conclusions, and in two appendices we collect various technical details
relevant to our calculations.

\section{Supersymmetry and Replicas}
\label{chiPT}

There are two different, yet equivalent, methods for performing
quenched chiral perturbation theory. The standard one is based on a
chiral Lagrangian extended to a supergroup \cite{BG}, and the other
on the replica method \cite{DS}. In the former, internal quark loops are
cancelled by a mechanism similar to the cancellation of vacuum diagrams
in space-time supersymmetric field theories. In the latter, quark loops
are removed by setting the number of quarks equal to zero.
The equivalence of the two methods in chiral perturbation theory
follows quite easily
once one sees the two different sets of Feynman rules. In both
cases quenching simply corresponds to removing the fermion determinant
in the original theory.

\noi
Let us begin at the quark level. Because we shall be interested in
computing quantities that include  quarks on external
lines, we denote by $N_v$ the number of such ``valence quarks''. In the
supersymmetric formulation we include appropriate sources $J$ for these
valence quarks (for simplicity restricted to quark bilinears here), but
cancel their contributions, at vanishing sources, by quarks of wrong
statistics. The generating functional is thus
\beq
Z_{Susy}[J] ~=~ \int\![dA_{\mu}]\frac{\det(i\slash{D}-m_v+J)}
{\det(i\slash{D}-m_v)^{N_{v}}}
e^{-S[A]} ~, \label{Suzy}
\eeq
where the determinant in the numerator is over $N_v$ fields.
The resulting theory is
not space-time supersymmetric, because there are no superpartners of
the gluons, but internal quark loops are cancelled by their ghost partners.

\noi
In the replica method we embed the $N_v$ valence quarks in a theory
with $N$ quarks in total (all of which are of ordinary fermionic statistics).
At that stage the generating functional reads
\beq
Z_{Replica}[J] ~=~ \int\![dA_{\mu}]\det(i\slash{D}-m_v+J)
\det(i\slash{D}-m_v)^{N-N_{v}}e^{-S[A]} ~,
\eeq
where the first determinant is taken over $N_v$ fields. The dependence
on $N$ is parametric, and the limit $N\to 0$ can be taken:
\beq
Z_{Replica}[J] ~=~ \int\![dA_{\mu}]\frac{\det(i\slash{D}-m_v+J)}
{\det(i\slash{D}-m_v)^{N_{v}}}
e^{-S[A]} ~. \label{rep}
\eeq
It is not just that the two methods succeed in removing the quark
determinant in the partition function. The two {\em generating functionals}
(\ref{Suzy}) and (\ref{rep}) are simply identical. Passing to the effective
field theory description, the equivalence is thus guaranteed from the very
beginning. This can easily be verified at the perturbative level of the chiral
Lagrangian with the same sources \cite{DS}. It is also clear that
it trivially generalizes to the partially quenched case.

\noi
We assume that in the quenched theories so constructed, the
chiral flavour symmetry is spontaneously broken to the diagonal subgroup. The
low energy degrees of freedom are the corresponding Goldstone bosons,
the dynamics of which can be described by a chiral Lagrangian.
In the replica method, the symmetry breaking pattern is as in QCD $SU(N)_L
\times SU(N)_R \rightarrow SU(N)_{L+R}$, while in the supersymmetric case
we have naively a graded flavour symmetry of the breaking pattern
$SU(N_v|N_v)_L \times  SU(N_v|N_v)_R \rightarrow SU(N_v|N_v)_{L+R}$.
One new feature common to both methods comes from the
fact that in contrast with the full theory the flavour singlet
cannot be integrated out: it is a degree of freedom that does not
decouple. This is well-known in the supersymmetric formulation, and
it is even more easily understood in the replica formalism. There is
simply no replica limit $N\to 0$ of a theory with $SU(N)$ symmetry,
and the trace part must be allowed to
fluctuate.\footnote{There is no corresponding problem with partial
quenching, i.e. taking the replica limit $N\to 0$ of a partially quenched
theory of flavour group $SU(N_f+N)$, with $N_f \geq 2$. Then the $m_0$-term is
not required to define the theory, and it can be decoupled.}

\noi
Although
the perturbative series obtained by the two
methods coincide, we need to go beyond perturbation theory here,
as we shall be treating the zero momentum modes of the Goldstone bosons
exactly. Then a very precise definition of the supersymmetric Haar
measure in the supersymmetric formulation, different from the one
employed in formal perturbative expansions, is required.
This has been explained in detail in refs. \cite{Z,DOTV} (see also the recent
discussion \cite{SS}). To obtain the precise Haar measure over which
to integrate, a more careful analysis of the flavour symmetries with bosonic
and fermionic quark species is required. The outcome is that one should
replace the naive supersymmetric generalization $U(N_v|N_v)$ by $Gl(N_v|N_v)$,
or rather what in the mathematics literature is called the maximally
symmetric Riemannian submanifold \cite{Z} of this supergroup (for simplicity
of notation we will not make
this distinction in what follows, and will just denote it by
$Gl(N_v|N_v)$). It basically amounts to choosing well-defined integration
paths for the fields involved; in particular assuring that
the action is bounded from below in all bosonic directions.
The replica method is in principle much easier since
no supergroup is needed at all, and one is working with an ordinary chiral
Lagrangian throughout. The perturbative Feynman rules are simpler, but
at the non-perturbative level it is not known how to go beyond series
expansions in general \cite{DS1} (an exception is QCD$_3$, where the
expansion terminates \cite{ADDV} and it thus gives the exact result
\cite{Szabo}). So while equivalence is trivial in
perturbation theory, it is highly non-trivial at the
non-perturbative level. By doing all computations both ways, we thus get
important cross-checks on our results. 

\section{Regimes of (Quenched) Chiral Perturbation Theory}
\label{regimes}

Let us consider QCD in a toroidal volume $V$ of average length scale
$L = V^{1/4}$. We assume that the volume is large with respect to the QCD
scale.
As in an infinite volume the lightest degrees of freedom are the Goldstone
bosons
of chiral symmetry breaking. They are describable in terms of a
chiral Lagrangian given by an
expansion in powers of the pion momentum $p = 2 \pi n/L$ and mass
$m_\pi$ over the cutoff of the effective theory, $\Lambda \simeq 4 \pi F$.
This is the standard chiral expansion, in which both quantities $p$ and
$m_\pi$ are taken to be of the same order.
At leading order, the chiral Lagrangian in a $\theta$ vaccuum is usually writen as
\beq
{\cal{ L }}^{(2)} =  {F^2\over 4}  {\rm Tr} \left(\partial_\mu U
\partial_\mu U^\dagger \right) - {\Sigma \over 2} {\rm Tr}
\left( {\cal M} e^{i\theta/N_f} U +  e^{-i\theta/N_f} U^\dagger  {\cal M}^{\dagger}\right) +
O\left(\frac{p}{\Lambda}\right)^4\ ,
\eeq
where $\Sigma$ and $F$ are the infinite volume quark condensate and
pion decay constant, both in the chiral limit. ${\cal M}$
is the quark mass matrix that for
simplicity we will take as proportional to the identity, ${\cal M}= m I$.
Furthermore, $U =
e^{i \sqrt{2} \xi/F}$ is an element of $SU(N_f)$. In the chiral expansion
$U$ is expanded around the classical solution $U = 1 + i \sqrt{2}
\xi/F+...$ Consider
the quadratic term in the action in momentum space
\beq
{\cal{ S }}^{(2)} =  \frac{V}{2} (p^2+m_\pi^2)  {\rm Tr}[ \xi_p^2]\ .
\label{eq:quadratic}
\eeq
It provides a gaussian damping factor that limits the fluctuations of
$Tr[\xi^2_p]$ to be
$$
O\left(\frac{1}{V (p^2 + m_\pi^2)}\right)
\sim O\left(\frac{1}{V m_{\pi}^{2}}\right)\ .
$$
When the linear size $L$ is much larger than the
Compton wavelength of the pions
$1/m_{\pi}$, the system hardly feels the finite volume, and the typical
momentum scale is thus
$p \sim m_{\pi}$.
So for $m_{\pi} \gg 1/L$, the fluctuations
are much smaller than $1/L^2$ and the expansion around the classical solution
is an expansion in powers of a quantity much smaller
than $1/(LF)^2 \ll 1$. As the size of the box or the quark mass
becomes smaller, finite size effects start to become important, but provided
$m_\pi L \geq 1$, ordinary perturbation theory is still applicable. In the
boundary of this regime, when  $m_\pi L \sim 1$, the expansion of the field
$U(x)$ around the
classical solution and the expansion in powers of momenta of the Lagrangian
itself become the same
expansion in powers of $1/F L$. This is the so-called
{\it $p$-expansion} \cite{GL} in which
$$
\frac{|\xi_p|}{F}~\sim~\frac{m_\pi}{\Lambda} ~\sim~ \frac{p}{\Lambda}
~\sim~ \frac{1}{L F}\ .
$$
If the chiral limit is approached further in such a way that the Compton
wavelength of the pion is larger than the box
size ($L> 1/m_\pi$), 
the conventional $p$-expansion eventually breaks down due to
propagation of pions with zero momenta \cite{GL}. Indeed, according to
eq.~(\ref{eq:quadratic}), when $m_\pi L^2 \sim 1$ (in units of the cut-off,
i.e. $m_{\pi} L^2 F = (2m\Sigma V)^{1/2} \sim 1$) the fluctuations of
the zero modes, $\xi^2_0$, are of $O(1)$
and the perturbative expansion for these modes breaks
down.
But the non-zero modes still have fluctuations that are
$$
O\left(\frac{1}{V p^{2}}\right) ~<~ O\left(\frac{1}{L^2}\right)
$$
and they are still perturbative. A convenient expansion for this
regime is the so-called {\it $\epsilon$-expansion}, in which \cite{GL}
$$
\frac{m_\pi}{\Lambda} ~\sim~ \frac{p^2}{\Lambda^2} ~\sim~
\frac{1}{L^2 F^2} ~\sim~ \epsilon^2 ~.
$$
The zero modes of the pion can be isolated by
factorizing $U(x)$ into a  constant
collective field $U_0$ and  the pion fluctuations $\xi (x)$:
\beqn
U(x) ~=~ U_0 \exp i \frac{\sqrt{2}\xi(x)}{F}\ .
\label{eq:fac}
\eeqn
The difficulty in this regime comes from the fact that
the integral over $U_0$ needs to be done exactly, while ordinary
chiral perturbation theory applies to the non-zero mode integration,
since $\xi_p^2 \sim O(\epsilon^2)$.
To leading order, the partition function is \footnote{The measure contains additional terms of order $\epsilon^2$ and higher
\cite{H} due to the change of variables to coordinates $\xi(x)$. These
terms are not needed to the order at which we will be working.}
\beqn
{\cal Z}(\theta,{\cal M}) = \int_{SU(N_f)} d\xi dU_0 \exp \left[  {1\over 2}
\int d^4 x  {\rm Tr} \left(\partial_\mu \xi
\partial_\mu \xi \right) + {m \Sigma V \over 2} {\rm Tr }
\left( e^{i\theta/N_f} U_0 + e^{-i \theta/N_f} U_0^{\dagger}\right)
\right].
\nonumber
\eeqn
If $F L \gg 1$, the $p$- and $\epsilon$-expansions should match in the range
of quark masses such that $m_\pi L \sim 1$. In this regime $m \Sigma V \sim
(F L)^2 (m_\pi L)^2 \gg 1$, so the results in the $\epsilon$-expansion
should reproduce those of the $p$-expansion in the limit of large $m \Sigma V$.

\noi
It is also interesting to consider averages in sectors of fixed
topology \cite{LS}.
Fourier-transforming in $\theta$, we get, to the same order:
\beqn
{\cal Z}_\nu({\cal M}) \!=\! \int_{SU(N_f)} d\xi \int_{U(N_f)}
\hspace{-.5cm} dU_0 \left({\rm det}
U_0\right)^\nu \exp\! \left[  {1\over 2} \int d^4 x {\rm Tr}
\left(\partial_\mu \xi\partial_\mu \xi \right) +
{m \Sigma V\over 2} {\rm Tr } \left( U_0 +
U_0^{\dagger}\right )\! \right]. \nonumber
\eeqn
The $\epsilon$- and $p$-expansions can be defined in the same way in this truncation of the theory.

\subsection{\sc Quenched Chiral Perturbation Theory}

\noi
In the supersymmetric formulation of the quenched theory,
assuming the pattern of chiral
symmetry breaking of the previous section, the low-energy behaviour of the
theory can be described by the supersymmetric chiral Lagrangian,
\beqn
{\cal L}^{(2)} ~=~\frac{F^2}{4} {\rm Str } \left(\partial_\mu U^{-1} \partial_\mu U \right) - {m \Sigma \over 2} {\rm Str }
 \!\!\left( U_{\theta} U + U^{-1} U^{-1}_{\theta}\right) +
{m_0^2 \over 2 N_c} \Phi^2_0 + \frac{\alpha}{2N_{c}}\partial_{\mu}
\Phi_0(x)\partial^{\mu}\Phi_0(x)\ ,
\label{Lrep}
\eeqn
where $\rm Str$ denotes the supertrace, $\Phi_0 \equiv \frac{F}{\sqrt{2}} {\rm Str}[-i \log(U)]$ and $U_\theta\equiv
\exp(i \theta/N_v) I_{N_v} + \tilde{I}_{N_v}$. Here, $I_{N_v}$
is the identity matrix in the fermion--fermion block of ``physical''
Goldstone bosons and zero otherwise, while
$\tilde{I}_{N_v}$ is the identity in the boson--boson block
and zero elsewhere.
As explained in the previous section, the integral over the Goldstone fields
is over a submanifold of $Gl(N_v|N_v)$.
An important difference with the full theory is
the unavoidable presence of the singlet field, which cannot be
decoupled in this case. Actually, the chiral expansion signals the breakdown
of the perturbative series if the singlet mass $m_0$ becomes comparable to
or larger than the cutoff scale in the effective theory $\Lambda \sim 4
\pi F$. To see this, let us consider the case of $N_v=1$ in the context
of the supersymmetric method. Using the parametrization of the Goldstone
submanifold of $Gl(1|1)$ from ref. \cite{DOTV},

\begin{center}
$U(x)$ = \( \left( \begin{array}{cc} e^{i\phi(x) \frac{\sqrt{2}}{F}} & 0\\
             0  & e^{s(x) \frac{\sqrt{2}}{F}}  \end{array} \right)\)
$\exp$\(\left( \begin{array}{cc} 0 & \gamma(x) \frac{\sqrt{2}}{F}\\
             \beta(x)\frac{\sqrt{2}}{F}  & 0 \end{array}\right)\) \ ,
\end{center}
where $\phi(x)$ and $s(x)$ are the real bosonic fields,
and $\gamma(x)$ and $\beta(x)$ are the fermionic ones,
we consider the usual $p$-expansion about the classical
vacuum $U=1$. 
The action expanded to quadratic order is
\begin{eqnarray}
{\cal{ S }}^{(2)} & = &  \int d^4 x
 \left\{ \frac{1}{2} \left(\phi \left(-\partial^2 + \frac{2 \Sigma m}{F^2}\right)
\phi + s \left( -\partial^2 +\frac{2 \Sigma m}{F^2}\right) s +
(\phi + i s) \left(\frac{m_0^2-\alpha \partial^2}{N_c}\right) (\phi + i s)\right) \right. \nonumber\\
& + & \left. \gamma\left(-\partial^2+
\frac{2 \Sigma m}{F^2}\right)\beta ~\right\}.
\label{superquadratic}
\end{eqnarray}
where $m_{\pi}^2F^2 = 2m\Sigma$. 
In contrast to the naive choice of $U(1|1)$ as the coset
of chiral symmetry breaking, the gaussian integrals are absolutely convergent
for all momenta such that $(m_0^2 + \alpha p^2)/N_c < p^2 + m_\pi^2$ and to any 
finite order in $(m_0^2 + \alpha p^2)/N_c$ even if this condition is not 
satisfied. It is only then that we are able to make
meaningful statements about the magnitude of fluctuations.

Using the properly normalized measure we get to this order:
\beq
{\cal Z} ~=~ \frac{\det(-\partial^2+m_{\pi}^2)}{
\det(-\partial^2+m_{\pi}^2)^{1/2} \det(-\partial^2+m_{\pi}^2)^{1/2} } ~=~ 1 ~,
\eeq
and ${\cal Z}$ is thus
independent of $m_0^2$ in the absence of sources.
This explains, to this order, the puzzle of how the effective partition
function apparently could depend on one new parameter $m_0$, when, in
the absence of sources, it should equal unity by construction. Only when
we include external sources for the quarks can a dependence on $m_0$ appear;
 in particular individual propagators will depend on $m_0$. It follows
that $m_0$ must have at its origin gluonic dynamics, which is probed by
appropriate quark sources. This picture is in nice agreement with what
is found at the level of the zero modes, see below.

\noi
The fluctuations of the Fourier modes
of $s$ and $\phi$ are seen to be
\beqn
\langle \phi_p^2 \rangle = \frac{(1-\alpha/N_c) p^2 + m_\pi^2 - m_0^2/N_c}
{V (p^2+m_\pi^2)^2} , \;\;~~~
\langle s_p^2 \rangle = \frac{(1+\alpha/N_c) p^2 + m_\pi^2 + m_0^2/N_c}{V (p^2+m_\pi^2)^2}\ .
\label{eq:gaussquenched}
\eeqn
Clearly if $m^2_0/N_c$ is larger than the chiral scales $p^2$ or $m^2_\pi$,
the fluctuations of the fields are controlled by this parameter and the
perturbative expansion breaks down for all modes (i.e. zero and non-zero) when
$m^2_0/N_c \geq F^2$ \footnote{We assume the parameter $\alpha \sim 1$.}. This is an artefact of the quenched approximation,
related to the fact that the singlet cannot be integrated out.
Defining $\epsilon'^2 = (m_0^2)/(N_c F^2)$ and $\epsilon^2 = 1/(F L)^2$,
in the regime of the $p$-expansion we have
\beqn
\frac{1}{(F L)^2} \sim \frac{m_\pi^2}{\Lambda^2} \sim O(\epsilon^2),\;
\;\;~~ \frac{\phi^2_p}{F^2} \sim O(\epsilon^2-\epsilon'^2), \;\;~~
\frac{s^2_p}{F^2} \sim O(\epsilon^2+\epsilon'^2)\ .
\label{pexp}
\eeqn
In the case of the zero modes, eqs.~(\ref{eq:gaussquenched}) imply that the
series breaks down
when $m^2_\pi/\Lambda^2 \leq \epsilon' \epsilon^2$. For $\epsilon' > \epsilon^2$
this will happen before zero modes need to be treated separately
in the unquenched theory.  This is, however, not the case if we only consider
averages in sectors of fixed topology at finite $\nu$.
Using left and right invariance of the super-Haar measure,
the factorization into zero modes and non-zero modes goes through as in
the unquenched case eq.~(\ref{eq:fac}), and the partition function
restricted to fixed topology $\nu$ is to the order we need it
\beqn
\hspace{-0.2cm}{\cal Z}_\nu({\cal M}) ~=~ \frac{1}{\sqrt{2\pi\langle \nu^2\rangle}}
e^{-\nu^2/2\langle\nu^2\rangle}\;\;
 \int_{Gl(N_f|N_f)}\!\!  dU_0 d\xi  \left({\rm Sdet}U_0\right)^\nu
\!\exp\!\left[\! {m \Sigma V\over 2} {\rm Str} \!\!\left( U_0 + U_0^{-1}
\right) \right. \nonumber\\
\left. + \int\!\! d^4x  \!\left(
-\frac{1}{2} {\rm Str } \left(\partial_\mu \xi \partial_\mu \xi \right) -
{m_0^2 \over 2 N_c} {\rm Str}(\xi)^2 - \frac{\alpha}{2 N_c} \left( \partial_\mu {\rm Str}[\xi] \right)^2 \right)  +...\right], \nonumber
\label{superfixednu}
\eeqn
where ${\rm Sdet}$ is the superdeterminant. The distribution of winding
numbers is Gaussian with mean
\beq
\langle\nu^2\rangle ~=~ \frac{F^2m_{0}^2 V}{2N_{c}} ~,
\label{nusquare}
\eeq
a relation that will play a crucial role when we match results in the
$\epsilon$-regime with those of the $p$-expansion regime.
The otherwise menacing $\Phi_0$ has disappeared in the zero-mode
sector at fixed topology, leaving as its only trace the average distribution
of topological charges. As a result of this, the usual $\epsilon$-regime, where
\beqn
\frac{1}{(F L)^2} \sim \frac{m_\pi}{\Lambda} \sim O(\epsilon^2)\ ,
\eeqn
can be approached.
This result may seem puzzling, as it indicates that a more chiral regime can
be reached in sectors of fixed gauge field topology,
whereas after summing over topology we find that the double-pole term
of the quenched propagator blows up at a scale much before the regime
of the $\epsilon$-expansion is reached. In fact this is correct, and
the resolution is found by noting that also the quenched $\epsilon$-expansion
in sectors of fixed topology eventually fails, when
$|\nu|\to \infty$.
In the sum over topology,
the dominant contributions are those around $|\nu| \sim Fm_0 \sqrt{V}$,
and for topological charges that large, the $\epsilon$-expansion
breaks down even in sectors of fixed topology.
In other words, the sickness that we found in the direct
analysis for $m_\pi^2/\Lambda^2 < \epsilon' \epsilon^2$ reappears in this case
when summing over topology. Actually this problem of the perturbative
expansion close to the chiral limit is probably unresolvable, since it is
also found at the quark level: in a background gauge field with topological
charge $\nu$, the contribution of the topological zero
modes to any quark propagator is $|\nu|/m V$. Using eq.~(\ref{nusquare})
the average over topology implies that this contribution becomes of order
$Fm_0/(m \sqrt{V})$, which is of $O(1)$ when
$m \sim \epsilon' \epsilon^2$. 
Concerning the non-zero modes, the perturbative expansion
remains a simultaneous expansion in $\epsilon$ and $\epsilon'$ also at fixed
topology. We will
be working to lowest non-trivial order in {\it both} expansion parameters.

\noi
Let us now summarize the relevant Feynman rules. We have already displayed the
usual supersymmetric version of the Lagrangian and the propagators for
the bosonic fields in eq. (\ref{eq:gaussquenched}) for the $N_v=1$ case. The propagator for
the fermionic fields $\gamma$ and $\beta$ is, after a simple rescaling,
read off from eq.
(\ref{superquadratic}) to be simply a conventional bosonic propagator. These
rules differ by some signs from the usual ones based on $U(1|1)$ \cite{BG}.
These differences are irrelevant in perturbation theory.
 The generalization to $N_v > 1$ follows
similarly \cite{BG}, with a few sign changes. Let us introduce some
convenient notation and define
\beq
\Delta(x) ~\equiv~ \frac{1}{V}\sum_p \frac{e^{ipx}}{p^2+m_{\pi}^2}
~~,~~~~~ \bar{\Delta}(x) ~\equiv~ \frac{1}{V}\sum_p~\!'
~\frac{e^{ipx}}{p^2} ~,
\label{delta}
\eeq
where a prime on the sum indicates that zero momentum is excluded.
We will also need the
corresponding expressions for the peculiar double-pole term. Let
\beq
G(x) ~\equiv~ \frac{1}{V}\sum_p \frac{e^{ipx}}{(p^2+m_{\pi}^2)^2}
~~{\rm and}~~~~~ \bar{G}(x) ~\equiv~ \frac{1}{V}\sum_p~\!'
~\frac{e^{ipx}}{p^4}\ .
\label{g}
\eeq
One notes that
\beq
\int\! d^4x ~\bar{\Delta}(x) ~=~ \int\! d^4x ~\bar{G}(x) ~=~ 0 ~,
\label{barvanish}
\eeq
properties that greatly facilitate comparison with integrated
Ward identities in the $\epsilon$-expansion. There is also the
relation
\beq
\Delta(x) ~=~ \frac{1}{m_{\pi}^2V} + \bar{\Delta}(x) - m_{\pi}^2
\bar{G}(x) + \ldots ~, \label{Taylor}
\eeq
so that both $\bar{\Delta}(x)$ and $\bar{G}(x)$ immediately follow
from Taylor expanding the finite-volume massive pion propagator $\Delta(x)$.

\noi
In the replica formulation \cite{DS}, all of the above discussion also applies.
The action is that of eq.~({\ref{Lrep}) after replacing
${\rm Str}$ by ${\rm Tr}$ everywhere, and after
changing the group manifold to $U(N)$.
 The  $m_0$ term actually serves to
interpolate between $SU(N)$ and $U(N)$: when $m_0\to \infty$ we go from
$U(N)$ to $SU(N)$, but then there can be no replica limit.
The way the $m_0$ term serves  to allow a replica limit becomes
particularly transparent when one looks at the Feynman rules: in a quark
basis the propagator of the off-diagonal mesons will have the usual form
of $\Delta(x)$,
while the diagonal combination has a propagator (with $E$ being a $N\times N$
matrix with unity in every entry):
\beq
{\cal G}_{ij}(x) ~=~ \frac{1}{V}\sum_p\left[
\frac{\delta_{ij}}{(p^2+m_{\pi}^2)}- E_{ij}\frac{(m_0^2+\alpha
  p^2)/N_c}{(p^2+m_{\pi}^2)^2{\cal F}(p^2)}\right]e^{ipx} ~.
\label{Gij}
\eeq
Here
\beq
{\cal F}(p^2)~\equiv~1+\frac{m_0^2+\alpha p^2}{N_c}\left(
  \frac{N}{p^2+m_{\pi}^2}\right) ~.
\label{F}
\eeq
For any finite $N$ we can take the limit $m_0 \to \infty$,
in which case the double pole in the last
term is cancelled. We then get the ordinary propagator with a factor
of $1/N$ in front, combining with the first term to give the usual diagonal
propagator in the quark basis.
The singularity at $N=0$ is regularized by keeping
a finite $m_0$. The replica limit $N\to 0$ is then taken only at the
end of the calculation, keeping $m_0$ finite.


\section{The Chiral Condensate}
\label{chicondensate}

The leading order result for the chiral condensate in a finite volume and
fixed topological sector was computed from the fully and partially quenched chiral
Lagrangian in ref.~\cite{DOTV}. For full QCD the first correction to this
result follows directly from the calculation of Gasser and Leutwyler
\cite{GL}. It is also of interest to find this first correction in the fully
quenched theory, and this was recently done using the replica method
\cite{D01}. In this section we show how the same result is obtained from the
supersymmetric method.

\noi
Let us consider the simplest case of $N_v=1$, because the result clearly does
not depend on $N_v$. For the group $Gl(1|1)$, the chiral condensate in a
sector with fixed topology can  be defined as:
\beq
\Sigma_\nu (m) ~=~
 {1 \over V}{\partial \over \partial J}
 \ln {\cal Z}_\nu ({\cal M}_J) \left. \right|_{J=0}\ .
\label{sigma}
\eeq
where ${\cal M}_J = {\rm diag } (m+j, m)$.
At leading order in the $\epsilon$-expansion one finds ~\cite{DOTV,OTV}
\beqn
{\Sigma_\nu (\mu)\over \Sigma } ~=~
\mu \left( I_\nu (\mu) K_\nu (\mu) + I_{\nu +1} (\mu)
K_{\nu -1}(\mu) \right) +{\nu \over \mu} \ ,\label{zerocon}
\eeqn
where
 $I_\nu (\mu), K_\nu (\mu)$ are modified Bessel functions, and
$\mu = m\Sigma V$.

\noi
To derive the first correction to this result in the $\epsilon$-expansion
it is convenient to first calculate the 1-loop improvement of the
action due to the fluctuations of the non-zero momentum modes. To
$O(\epsilon^2)$ the contribution from the measure does not
affect the result, and to that order we simply evaluate
$$
\left\langle 1 - {\Sigma \over 2 F^2} {\rm Str }\left[ {\cal M}_J
\left( U_0 + U_0^{-1}\right)
  \int d^4x \xi^2(x) \right] \right\rangle \ .
$$
Performing the integral over $\xi$ to leading order,
and then re-exponentiating the correction, we obtain
\beqn
{\cal Z}_\nu({\cal M}_J) & = &
 \int_{Gl(1|1 )}\!\! dU_0  \left({\rm Sdet}U_0\right)^\nu \!\exp\!\left[\!
    {\Sigma_{eff} V\over 2}
  {\rm Str }\left({\cal M}_J \left(  U_0 + U_0^{-1}\right) \right)\right] ~.
\label{znu1loop}
\eeqn
Then the effective coefficient determining the strength of the condensate
at finite volume is
\beqn
\Sigma_{eff}(V) ~\equiv~ \Sigma
\left( 1 + {1 \over N_c F^2} (m_0^2 \bar{G}(0) +
\alpha {\bar\Delta}(0))\right) ~, 
\label{sigmaeff}
\eeqn
were $\bar{G}, \bar{\Delta}$ are defined in eqs.~(\ref{delta}-\ref{g}). 
The $m_0^2$ and $\alpha$ terms are what remains from the partial
cancellation of the fermionic propagator and the bosonic ones. The partition
function is then the same as the one at leading order with the change
$\Sigma \rightarrow \Sigma_{eff}(V)$.


\noi
It is now easy to calculate the condensate to one loop, by differentiating
eq.~(\ref{znu1loop}) as in eq.~(\ref{sigma}) to get
\beqn
\Sigma_\nu^{\rm 1-loop}(\mu)  &=& \Sigma_\nu(\mu')  {\mu' \over \mu} \cr
&=&
\Sigma_\nu (\mu) + 2 {1
 \over N_c F^2}  ( m_0^2 {\bar G}(0) + \alpha \bar{\Delta}(0))
\mu I_\nu (\mu) K_\nu (\mu) + \dots \ \ \ ,\label{newcon}
\eeqn
where
\beq
\mu' ~\equiv~ m \Sigma_{eff} V ~=~ \mu \left( 1 +
{1 \over N_c F^2} ( m_0^2 {\bar G}(0) + \alpha \bar{\Delta}(0)) \right).
\label{mu'}
\eeq


\noi
One peculiarity of the quenched approximation is that the
$O(\epsilon^2)$
correction to the condensate is ultraviolet divergent even in dimensional
regularization because of the double pole propagator \cite{laurent,CP}. We find
\cite{HL}
\beqn
{\bar G}(0) ~=~  \beta_2 + {1\over 8 \pi^2} \left(\ln (L/L_0)  -
c_1 \right)\ , \;\;\; {\bar \Delta}(0) ~=~  - \frac{\beta_1}{L^2}, \label{fvlog}
\eeqn
where $\beta_1$ and $\beta_2$ are two of the universal ``shape coefficients''. It
depends on the shape of the box, and the precise value can, for any given volume,
be computed from the general expression given in ref. \cite{HL}.
In dimensional regularization the constant $c_1$ reads
\beq
c_1 ~=~ \frac{1}{d-4} - \frac{1}{2} \left(\Gamma'(1) + 1 + \ln(4\pi) \right)
+ O(d-4) ~,
\eeq
and $1/L_0$ is the ultraviolet subtraction point.
The divergent term in $c_1$ matches precisely the 1-loop counterterm needed
for the
condensate in the infinite-volume theory \cite{CP}. Below we present the numerical results
corresponding to the subtractions that are required by the
$\overline{MS}$ scheme.

\noi
Note that the one-loop contribution contains a quenched finite-volume
logarithm~\cite{D01}, which reflects the infrared sickness of the quenched
approximation. It is the finite-volume counterpart of the usual logarithmic
chiral divergence in infinite-volume quenched chiral perturbation theory.
The analysis is clearly restricted to domains where this divergence
does not yet overwhelm the tree-level result.

\noi
Several methods have been proposed and implemented to calculate the
coupling $\Sigma$ by finite-size scaling techniques.
One method is to take the discontinuity of the chiral condensate
(\ref{newcon}) across the imaginary $\mu$-axis to get the spectral
density of the Dirac operator eigenvalues in this regime \cite{DOTV}.
Since the density depends on the parameter $\Sigma$,
the leading tree-level expression for the density can be
used to extract $\Sigma$ in this way \cite{EHKN}.
Alternatively, the condensate may also be computed directly on the lattice
\cite{DEHN,HJL,DG,Has} and one can then fit its finite size and quark-mass
dependence to the prediction in eq.~(\ref{newcon}).
Both methods have been shown to work well
in extracting $\Sigma$ by neglecting the
contributions of eq.~(\ref{newcon}). With sufficiently high statistics,
one could
also aim at a  determination of the pion decay constant $F$ \cite{GL,D01}
from these corrections.
In the quenched case
the problem is that it brings in two new unknowns: $m_0$ and $\alpha$.
To estimate the typical size of the correction, we can compare
the 1-loop-improved chiral condensate at two different volumes, after subtracting the trivial topological zero mode
contributions and keeping
$\mu = m\Sigma V$ fixed. In this way we do not have to address the issue
of the $L_0$ scale dependence\footnote{In the direct computation of the
quark condensate UV divergences appear at the quark level which must be subtracted
before comparing to eq.~(\ref{ratio}) \cite{HJL}.}:
\beq
\frac{\Sigma^{1-loop}_{\nu,L_{1}}(\mu)-|\nu|/\mu}
{\Sigma^{1-loop}_{\nu,L_{2}}(\mu)-|\nu|/\mu}
~=~ 1 + \frac{1}{4\pi^2}\frac{1}{N_cF^2}\frac{\mu I_{\nu}(\mu)
K_{\nu}(\mu)}{\Sigma_{\nu}(\mu)-|\nu|/\mu}\left(m_0^2 \ln\left(\frac{L_1}{L_2}\right) -\alpha
\beta_1
\left(\frac{1}{L_1^2} - \frac{1}{L_2^2}\right) \right)\ .
\label{ratio}
\eeq
As an example, for the smallest and largest lattice volumes considered in \cite{HJL} in the
$\nu=1$ sector, the second
term in eq.~(\ref{ratio}) is around $15\%$ for $m_0\sim 600$ MeV. Although the
 statistical errors in \cite{BBetal} are larger than this, this is not clearly
beyond hope in more precise studies.



\section{Quenched Correlation Functions}
\label{QCF}

Meson correlation functions can be computed systematically in the
$\epsilon$-expansion. Here we present the results of the
$O(\epsilon^2)$ calculation for scalar and
pseudoscalar correlation functions in the quenched theory.
We have performed all computations in both the supersymmetric
and replica formulations, checking that the final results are identical.
We begin with the flavour-singlet combinations involving
\beq
S^0(x) ~\equiv~ \bar{\psi}(x) I_{N_v}\psi(x)~,~~~~~~~
P^0(x) ~\equiv~ \bar{\psi}(x) i\gamma_5I_{N_v}\psi(x) ~,
\label{s0p0}
\eeq
where $I_{N_v}$ is the flavour
projector onto the physical quark sector or that of the
$N_v$ valence quarks. For simplicity we consider $N_v=1$ for the singlet
case.
To compute these correlation functions, we make use of the correspondence
between local sources added at the quark level
\beq
{\cal L}_{QCD} ~\to~ {\cal L}_{QCD} + s(x)S^0(x) + p(x)P^0(x) ~,
\eeq
and the substitution
\beq
{\cal M} ~\to~ \chi(x) = {\cal M} + s(x) I_{N_v} + ip(x) I_{N_v}~
\eeq
in the effective theory.
Two functional derivatives of the generating functional with respect
to $s(x)$ and $p(x)$ thus give us the scalar and the pseudoscalar
2-point functions, respectively, to the desired order. In the supersymmetric
formulation we follow the procedure outlined in section 2, and thus add
the sources in the fermion--fermion block only. Using the two sets
of Feynman rules for the fluctuation fields $\xi(x)$, we find, to
order $\epsilon^2$:
\beqn
\langle S^0(x)S^0(0) \rangle &=& C_S^0 +
\frac{\Sigma^2}{2 F^2} \left[\frac{1}{N_{c}} \left(m_0^2{\bar{G}}(x)
+ \alpha\bar{\Delta}(x)\right) a_- - \bar{\Delta}(x)
{a_+ + a_- -  4 \over 2}  \right]    \label{QS} \\
\langle P^0(x)P^0(0) \rangle &=& C_P^0 -
\frac{\Sigma^2}
{2 F^2} \left[ \frac{1}{N_{c}} \left(m_0^2{\bar{G}}(x)
+ \alpha\bar{\Delta}(x)\right) a_+ - \bar{\Delta}(x) {a_+ + a_- +
4 \over 2}\right] \ ,
 \label{QP}
\eeqn
with
\beqn
a_{\pm} ~=~ \langle(U_{11}\pm U^{-1}_{11})^2\rangle ~.
\label{apm}
\eeqn
The constant terms are given by the same expectation values of eq.(\ref{apm})
\beq
C_S^0 ~=~ \frac{\Sigma_{eff}^2}{4} a^{1-loop}_+ ~,~~~~~~
C_P^0 ~=~ -\frac{\Sigma_{eff}^2}{4} a^{1-loop}_- ~,
\eeq
but now evaluated with respect to the one-loop-improved action
(\ref{znu1loop}),
and $\Sigma_{eff}$ is as defined in eq. (\ref{sigmaeff}). In this
way we obtain these constant terms to the required order $\epsilon^2$.

\noi
These expressions are valid in both formulations. As a first check,
we note that when $U \to 1$ the $x$-dependent part of
$\langle S^0(x)S^0(0) \rangle$ vanishes: there is no tree-level
scalar propagation in the quenched theory. But there {\em should} be
tree-level propagation
of the quenched pseudoscalar singlet, and indeed  in this limit
the $x$-dependent part of $\langle P^0(x)P^0(0) \rangle$ approaches
\beq
\frac{2\Sigma^2}{F^2}\left[
\left(1 - \frac{\alpha}{N_c}\right) \bar{\Delta}(x) - \frac{m_0^2}{N_c} \bar{G}(x)
\right]
\eeq
which is simply the massless singlet propagator (cf. eq. (\ref{Gij})). We shall make a more detailed comparison,
also including the zero-momentum modes, below.

\noi
To evaluate the
remaining group integrals in closed form, we can make use of the exact
results obtained in the supersymmetric formulation \cite{DOTV}. For a source
$\mu_J \equiv \mu+J$ in the fermion--fermion slot,
the generating function becomes
\beq
{\cal Z}_{\nu}[J] ~=~
\frac{1}{2}\mu_J\left(I_{\nu+1}(\mu_J)+I_{\nu-1}(\mu_J)\right)K_{\nu}(\mu)
+ \frac{1}{2}\mu I_{\nu}(\mu_J)\left(K_{\nu+1}(\mu)+K_{\nu-1}(\mu)\right) ~.
\eeq
Using the fact that $Z_{\nu}[0]=1$ (which here is a consequence of two
Wronskian identities for the Bessel functions),
this gives us the expectation value
\beqn
\left\langle (U_{11}+(U^{-1})_{11})^2\right\rangle &=& \left.4
\frac{\delta^2}{\delta J^2}Z_{\nu}[J]\right|_{J=0} \cr
&=& 4\left[I_{\nu}(\mu)K_{\nu}(\mu)-I_{\nu+1}(\mu)K_{\nu-1}(\mu) +1
+ \frac{\nu(\nu-1)}{\mu^2}\right] ~.
\label{1111}
\eeqn
It is quite simple to find the remaining matrix element by means of the
explicit parametrization:
\beqn
\left\langle U_{11}(U^{-1})_{11}\right\rangle &=& \left\langle 1 +\gamma \beta
\right\rangle \nonumber\\
&=& 1 + 2\int_0^{2\pi}\!\frac{d\theta}{2\pi}\int_0^{\infty}\! ds d\gamma
d\beta ~\gamma\beta e^{\nu(i\theta-s)}e^{\mu(\cos(\theta)-\cosh(s))} \nonumber\\
&=& 1 + 2I_{\nu}(\mu)K_{\nu}(\mu) ~.
\eeqn
These expectation values suffice to compute the correlation functions
in eqs. (\ref{QS}) and (\ref{QP}).
Rewriting the final answer in terms of the tree-level
chiral condensate eq.~(\ref{zerocon}) and its derivative
\beq
\frac{\Sigma_{\nu}'(\mu)}{\Sigma} ~=~ I_{\nu}(\mu)K_{\nu}(\mu)
- I_{\nu+1}(\mu)K_{\nu-1}(\mu) - \frac{\nu}{\mu^2} ~, \label{Sigma'}
\eeq
we find:
\beq
a_+  =   4 \left[ \frac{\Sigma'_\nu(\mu)}{\Sigma} + 1 +
\frac{\nu^2}{\mu^2}\right] \ , a_- = 4
\left[-\frac{1}{\mu} \frac{\Sigma_\nu(\mu)}{\Sigma} +
\frac{\nu^2}{\mu^2}\right],
\eeq
while the constants to the same order are:
\beqn
C_S^0 &=& \Sigma_{eff}^2 \left[\frac{\Sigma'_{\nu}(\mu')}{\Sigma_{eff}} + 1 +
\frac{\nu^2}{\mu'^2}\right] ~=~ \Sigma^2 \left[\frac{\Sigma^{1-loop~'}_{\nu}
(\mu)}{\Sigma} + \left(\frac{\Sigma_{eff}}{\Sigma}\right)^2 +
\frac{\nu^2}{\mu^2}\right] \cr
C_P^0 &=& \Sigma^2_{eff}\left[\frac{1}{\mu'} \frac{\Sigma_{\nu}(\mu')}{
\Sigma_{eff}} - \frac{\nu^2}{\mu'^2}\right] ~=~ \Sigma^2\left[\frac{1}{\mu}
\frac{\Sigma^{1-loop}_{\nu}(\mu)}{\Sigma} - \frac{\nu^2}{\mu^2}\right] \ ,
\label{cpcss}
\eeqn
where $\mu'$ and $\Sigma_{eff}$ are defined in eqs.~(\ref{mu'}) and
(\ref{sigmaeff}), respectively.

\noi
Using the relation 
\beq
\frac{\Sigma_{\nu}'(\mu)}{\Sigma} + \frac{1}{\mu}
\frac{\Sigma_{\nu}(\mu)}{\Sigma} ~\to~ 0 ~~~~~{\mbox{\rm as}}~~\mu ~\to~ 0 ~,
\label{finitesum}
\eeq
it is easy to show that in the sum of  the scalar and pseudoscalar correlation functions, 
the poles in the quark mass due to the zero modes cancel,
and the sum has then a well-defined massless limit even
at finite volume for $\nu \neq 0$. The case $\nu=0$ is special: there is then an additional
infrared singularity due to the distribution of the smallest non-zero Dirac
operator eigenvalue. However,
the infrared divergence for $\mu \to 0$ in that case is only
logarithmic.

\noi
It is instructive to see, analytically, how the $p$- and $\epsilon$-expansions
match one another when $m_\pi L \sim 1$ (with $F L \gg 1$).
We have already noted above that the $x$-dependent
part of the $\langle P^0(x) P^0(0)\rangle$ correlation function to
$O(\epsilon^2)$ approaches the tree-level propagator of a massless
flavour singlet. Now that we have the constant part of this correlation
function evaluated explicitly, we can see how this constant part precisely
serves, to leading order, to restore the full massive propagator of the
$p$-expansion. To see the matching we need to sum over topology first,
because this is how the $p$-expansion is usually done. After performing
this sum over topology, using relation (\ref{nusquare}) as well as the
Gell-Mann--Oakes--Renner relation, we get, to leading order:
\beqn
\langle P^0(x)P^0(0)\rangle &\sim &
\Sigma^2\left(\frac{1}{\mu} - \frac{\langle\nu^2\rangle}{\mu^2}\right)
+ \frac{2\Sigma^2}{F^2}\left[\left(1-\frac{\alpha}{N_{c}}\right)
\bar{\Delta}(x) - \frac{m_{0}^2}{N_{c}}\bar{G}(x)\right] \cr
&=& \frac{2\Sigma^2}{F^2}\left[\frac{1}{m_{\pi}^2V} - \frac{m_{0}^2}{
N_{c}}\frac{1}{m_{\pi}^4V} +\left(1-\frac{\alpha}{N_{c}}\right)
\bar{\Delta}(x) - \frac{m_{0}^2}{N_{c}}\bar{G}(x)\right] \cr
&=& \frac{2\Sigma^2}{F^2}\left[\Delta(x) - \frac{\alpha}{N_{c}}
\sum_p \frac{p^2 e^{ipx}}{(p^2+m_{\pi}^2)^2} - \frac{m_{0}^2}{N_{c}}
G(x) + O(m_{\pi}^2)\right] \cr
&=& \frac{2\Sigma^2}{F^2}{\cal G}(x)
+ O(m_{\pi}^2) ~.
\eeqn
As in the case of the one-loop correction to the chiral condensate \cite{D01},
the sum over topology precisely restores the $m_0$-dependence needed for
the $\epsilon$- and $p$-expansions to match.

\noi
For comparisons with lattice gauge theory data, it is
convenient to recast the above results in terms of the space-integrated
correlators in the (euclidean) time direction, here labelled by $t$:
\beq
{\cal P}^i(t) = \int d^3 x \langle P^i(x) P^i(0) \rangle \;\;\;\; ,
~~~~ {\cal S}^i(t) = \int d^3 x \langle S^i(x) S^i(0) \rangle.
\eeq
For this purpose we need
the projections onto zero momentum of both the usual massless propagator
and the double-pole term. They both follow immediately from Taylor-expanding
 the massive pion propagator in powers of $m_\pi$. From
\beq
\int\! d^3x~ \Delta(x)
~=~ \frac{\cosh(m_{\pi}(T/2 - t))}{2m_{\pi}\sinh(m_{\pi}T/2)} ~,
\label{finiteVcor}
\eeq
and  eq. (\ref{Taylor}), we find
\beq
\int\! d^3x~ \Delta(x) ~=~ \frac{1}{Tm_{\pi}^2}
+ \frac{T}{2}\left[ \left(\tau - \frac{1}{2}\right)^2 - \frac{1}{12}\right]
+ \frac{T^3}{24}\left[\tau^2(\tau-1)^2-\frac{1}{30}\right]m_{\pi}^2 + \ldots ,
\eeq
which makes it convenient to define \cite{HL}
\beqn
h_1(\tau) &=& \frac{1}{T}\int\! d^3x~ \bar{\Delta}(x) ~=~
\frac{1}{2}\left[(\tau - \frac{1}{2})^2 - \frac{1}{12}\right] \\
h_2(\tau) &=& -\frac{1}{T^3}\int\! d^3x~ \bar{G}(x) ~=~
\frac{1}{24}\left[\tau^2(\tau-1)^2-\frac{1}{30}\right] ~.
\eeqn
Here $T$ is the extent in the temporal direction, and $\tau = t/T$. These
2nd and 4th order polynomials replace the exponentially decaying
correlations in the regime of the $\epsilon$-expansion.

\noi
Figures~\ref{fig:singlets} show the result for ${\cal P}^0(t)$ and
${\cal S}^0(t)$ in this $\epsilon$ regime, for three values of the
topological charge $\nu=0,1,2$,
and a quark mass $m=3$ MeV in a typical lattice of $V=16^3 32$ at
$a^{-1} = 2$ GeV. In these plots, and in those of the next sections,
we have used the values $m_0 = 600$ MeV and $\alpha=0.6$. For $\Sigma$ we
take as indicative the value ($270$ MeV$)^3$ at
the commonly used renormalization scale $L_0 = (4\pi F)^{-1}$.

\begin{figure}
\vspace{0.0cm}
\hspace{-0.0cm}
\epsfig{file=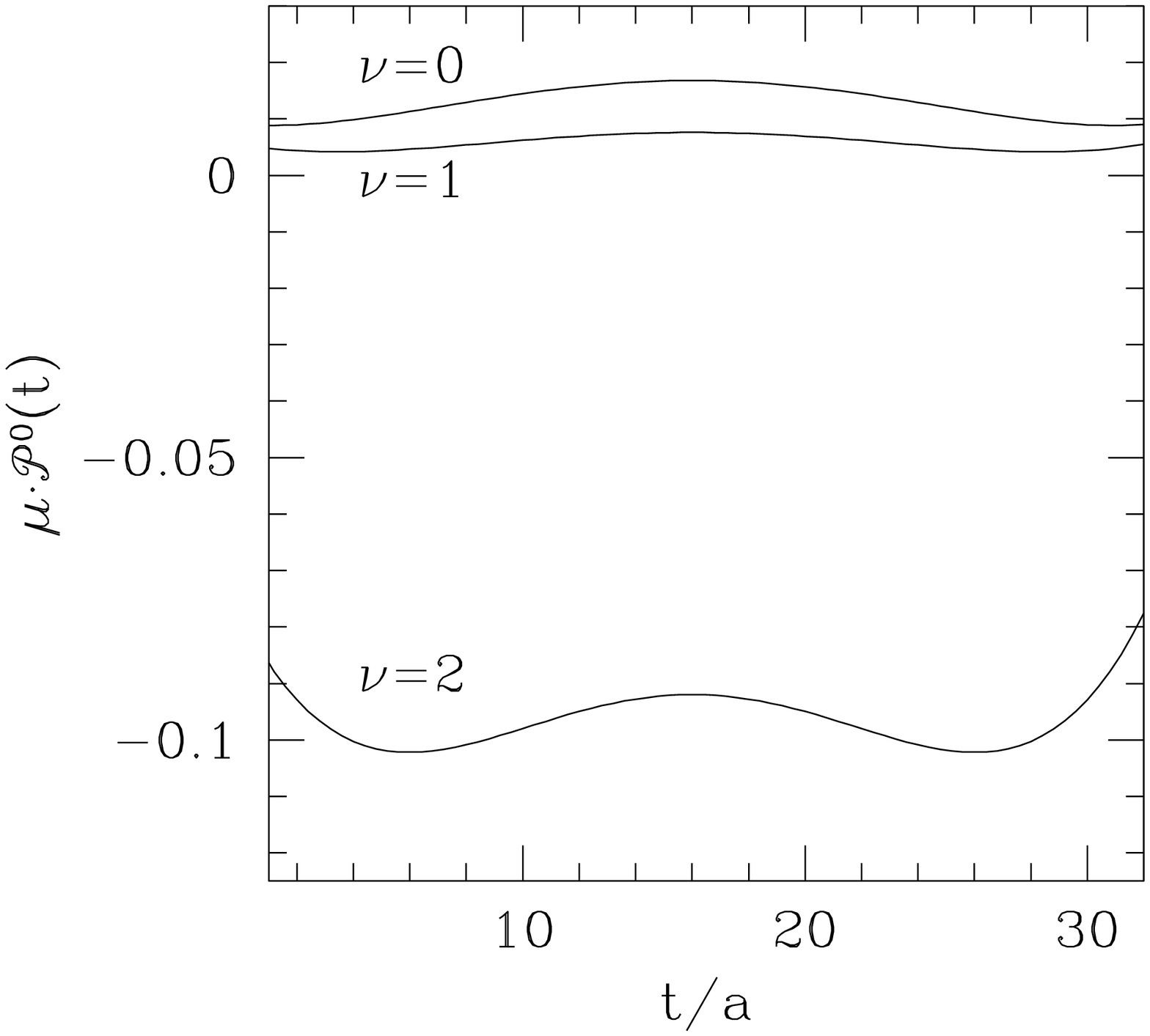,height=9cm,width=8cm}
\epsfig{file=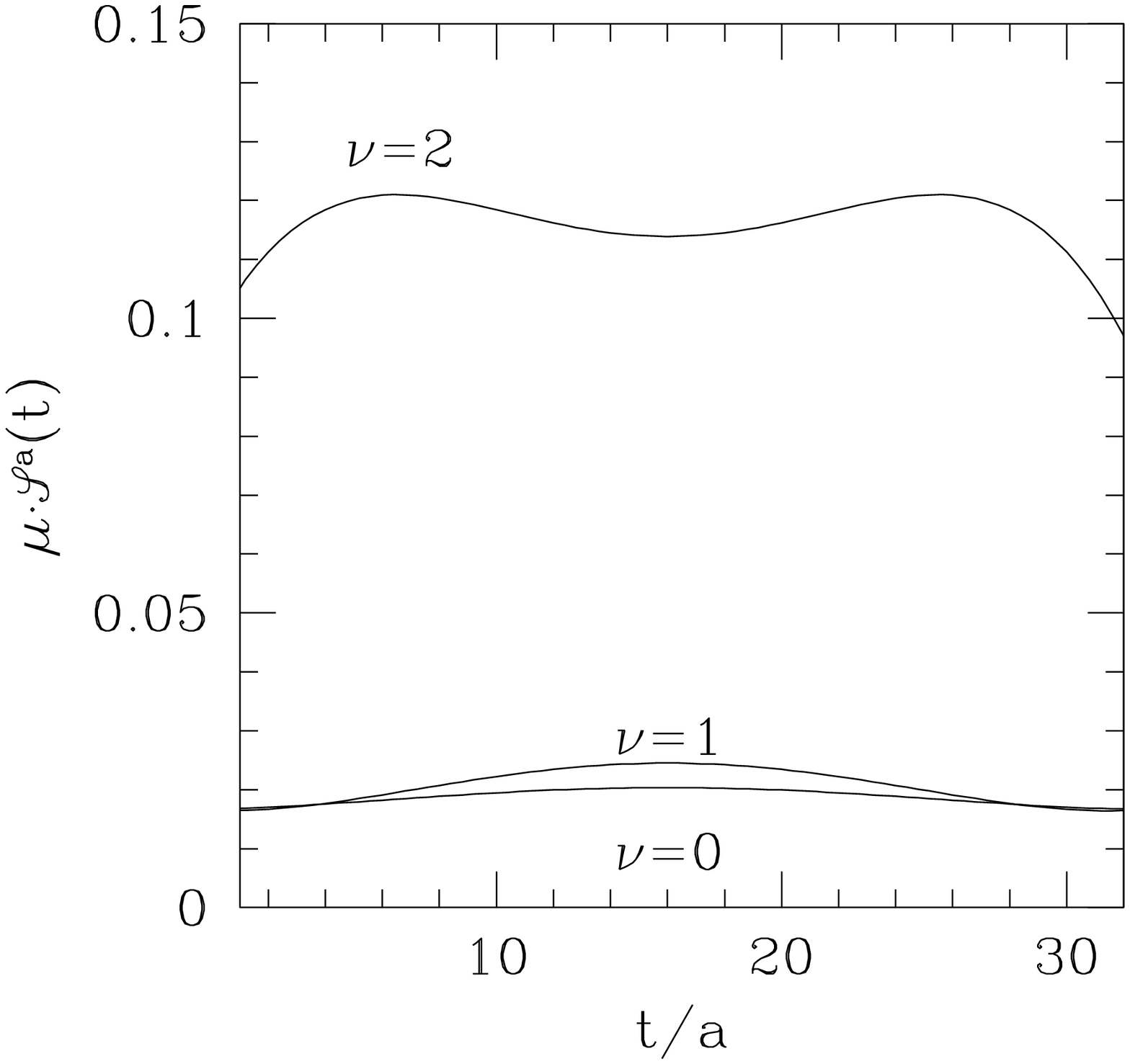,height=9cm,width=8cm}
\caption{ \label{fig:singlets} $\mu {\cal P}^0(t)$ (left) and $\mu {\cal S}^0(t)$ (right) in a lattice of
$V=L^3 T = 16^3 32$ and $a^{-1} \simeq 2$ GeV. The quark mass is $m=3$ MeV and
the three curves correspond to $\nu=0,1$ and 2.}
\end{figure}

\noi
We should stress that in these predictions, and in any other predictions given
in
this paper, the contributions from higher excited states with the same
quantum numbers are neglected. These higher
states clearly do contribute to the correlation functions at short time
separations,
but they are exponentially suppressed at large $t$.
Therefore our predictions should only fit
the measured correlation functions for sufficiently large time separation
between the sources.

\subsection{\sc Flavoured Correlation Functions}
\label{flavoured}

The computation of mesonic correlation functions in non-singlet channels
is a little more difficult, because we need the exact evaluation of
the group integral over zero momentum modes for $N_v > 1$.
Here we restrict ourselves to the simplest case of $N_v=2$.

\noi
Let us denote the quark bilinears as follows:
\beq
S^a(x) ~\equiv~ \bar{\psi}(x) t^a  I_{N_v} \psi(x)~~~~,~~~~~~~
P^a(x) ~\equiv~ \bar{\psi}(x)t^a i\gamma_5  I_{N_v} \psi(x) ~,
\eeq
where in this case $t^a = \tau^a/2$, and the $\tau^a$'s are the usual
Pauli matrices. The corresponding two-point functions are $\langle S^a(x)S^a(0)\rangle$ and $\langle P^a(x)P^a(0)\rangle$, for $a$=1-3. 
Here we present some details of the calculation of the correlators for
$a=1$.

\noi
The appropriate flavoured sources $s^a(x)$ and $p^a(x)$ are obtained by
lettting
\beq
{\cal L}_{QCD} ~\to~ {\cal L}_{QCD} + s^a(x)S^a(x) + p^a(x)P^a(x) ~
\eeq
at the quark level. This translates into
\beq
{\cal M} ~\to~ \chi(x) = {\cal M} + s^a(x)t^a I_{N_v} + ip^a(x)t^a I_{N_v} ~
\eeq
in the effective theory. Two functional derivatives with respect to these
sources then give us the desired correlation functions. To order $\epsilon^2$,
we find:
\beqn
\langle S^1(x)S^1(0) \rangle &=& C_S^1 +
\frac{\Sigma^2}{2 F^2} \left[\frac{1}{N_{c}} \left(m_0^2{\bar{G}}(x)
+ \alpha\bar{\Delta}(x)\right)
c_- - \bar{\Delta}(x) b_-  \right] \label{QS1} \\
\langle P^1(x)P^1(0) \rangle &=& C_P^1 -  \frac{\Sigma^2}{2 F^2} \left[
\frac{1}{N_{c}} \left(m_0^2{\bar{G}}(x)
+ \alpha\bar{\Delta}(x)\right)c_+ - \bar{\Delta}(x) b_+\right] \ ,
 \label{QP1}
\eeqn
where, after using some of the identities provided in Appendix \ref{gl22},
one finds
\beqn
c_{\pm} &=& \frac{1}{4} \left\langle (U_{12} + U_{21} \pm (U^{-1})_{12}
\pm (U^{-1})_{21})^2 \right\rangle \\
b_{\pm} &=& \frac{1}{2} \left\langle U_{11} U_{22} + (U^{-1})_{11}
(U^{-1})_{22} \right\rangle \pm 1 \ .
\eeqn

\noi
The constant terms will also be evaluated to
$O(\epsilon^2)$.
Since they are flavour-symmetric, we can simply take the
constant part of
$$
\frac{1}{3}\frac{\delta^2}{\delta s^a(x) \delta s^a(0)}{\cal Z}_{\nu}[J]
$$
and similarly for $C_P^a$. Using the $SU(2)$ completeness relation
\beq
t^a_{ij}t^a_{kl} ~=~ \frac{1}{2}\left[\delta_{il}\delta_{jk} - \frac{1}{2}
\delta_{ij}\delta_{kl}\right] \ ,
\eeq
we then get
\beqn
C_S^a &=& \frac{\Sigma_{eff}^2}{24}
\left\langle\frac{1}{2}\left((U_{11}+(U^{-1})_{11})^2
+ (U_{22}+(U^{-1})_{22})^2\right) + 2(U_{12}+(U^{-1})_{12})
(U_{21}+(U^{-1})_{21})\right. \cr &&
- \left.(U_{11}+(U^{-1})_{11})(U_{22}+(U^{-1})_{22})
\right\rangle \cr
C_P^a &=& -\frac{\Sigma_{eff}^2}{24}
\left\langle\frac{1}{2}\left((U_{11}-(U^{-1})_{11})^2
+ (U_{22}-(U^{-1})_{22})^2\right) + 2(U_{12}-(U^{-1})_{12})
(U_{21}-(U^{-1})_{21})\right. \cr
&& - \left.(U_{11}-(U^{-1})_{11})(U_{22}-(U^{-1})_{22})
\right\rangle ~.
\eeqn
Here the expectation values are again to be computed with
respect to the one-loop-improved action eq.~(\ref{znu1loop}) so as to reach
$O(\epsilon^2)$ accuracy.

\noi
In order to calculate the various integrals over the zero-momentum modes
explicitly, we make
use of the fact that they can all be related to an expectation value already
computed in eq. (\ref{1111}), and
\beq
    \left\langle(U_{11} + (U^{-1})_{11})(U_{22}+(U^{-1})_{22})\right\rangle \
    ,
\label{1122}
\eeq
which was evaluated explicitly for the zero-momentum mode integral with
$N_v=2$ in ref. \cite{TV}. The idea of how to get the other integrals from
these two is simple, and most easily explained
in the replica formalism. There it is particularly clear that the
quantity $\det({\cal M})^{-\nu}{\cal Z}_{\nu}$
is a function of the eigenvalues of
${\cal M}{\cal M}^\dagger$ only.
This means that all quadratic expectation values, after
appropriate manipulations, can be related directly to the expectation
value of eq.(\ref{1122}), and that of eq. (\ref{1111}).

\noi
Using a variety of   $2\times 2$
matrix sources $J$, we have collected a list with the results of explicit
evaluations of all needed zero-momentum mode integrals.
For the reader interested in the technical details, see Appendix \ref{gl22}. Here we simply present the final results:
\beqn
b_+ & = & 2 \left( 1 + \frac{\nu^2}{\mu^2}\right)\  ,\ \ \ \ \ b_- =
2 \frac{\nu^2}{\mu^2} \cr
c_+ & = & 2 \frac{\Sigma_\nu'(\mu)}{\Sigma}\  ,\ \ \ \ \
 c_- = - 2 \frac{1}{\mu} \frac{\Sigma_\nu(\mu)}{\Sigma} \;\;\;
\eeqn
and
\beq
C_S^a = \frac{1}{2}\Sigma\Sigma^{1-loop~'}_{\nu}(\mu)\ ,\ \ \ \ \
C_P^a = \frac{1}{2\mu}\Sigma\Sigma^{1-loop}_{\nu}(\mu)  ~.
\label{cpcsns}
\eeq

\noi
As a non-trivial
check on the calculation, we have also evaluated the correlators corresponding
to $a=2$ and 3. In the latter case the intermediate details are very
different, but the final results coincide in both cases
with the ones above for $a=1$, as expected.

\noi
In the sum of the two flavoured correlation functions, our expressions
simplify. We find (no sum on $a$):
\beqn
\!\!\!\!\!\!\langle S^a(x)S^a(0) \rangle\!\! &+&\!\!
\langle P^a(x)P^a(0) \rangle
~=~ \frac{1}{2}\Sigma\left[\Sigma^{1-loop~'}_{\nu}(\mu) +
\frac{1}{\mu}\Sigma^{1-loop}_{\nu}(\mu)\right] \cr
&& + \frac{\Sigma^2}{F^2}\left[\bar{\Delta}(x) -
\left(\frac{\Sigma_{\nu}(\mu)}{\Sigma} + \frac{1}{\mu}
\frac{\Sigma_{\nu}(\mu)}{\Sigma}\right)\frac{1}{N_{c}} \left(m_0^2{\bar{G}}(x)
+ \alpha\bar{\Delta}(x)\right)\right] ~.
\label{SaPasum}
\eeqn
As in the flavour-singlet case, the sum remains finite
in the $\mu \to 0$ limit for $\nu \neq 0$, on account of eq. (\ref{finitesum}).

\noi
Here we can also check analytically that the $\epsilon$- and $p$-expansions
join on to each other. This is because the $\langle P^a(x)P^b(0)\rangle$
correlation function we have computed should simply reproduce the tree-level
pion propagation of the $p$-expansion to $O(\epsilon^2)$. Indeed, we
find, to lowest order (no sum on $a$):
\beqn
\langle P^a(x)P^a(0)\rangle & \sim & \left[\frac{\Sigma^2}{2\mu}
+ \frac{\Sigma^2}{F^2}\bar{\Delta}(x)\right] \cr
&=&
\frac{\Sigma^2}{F^2}\left[\frac{1}{m_{\pi}^2V} + \bar{\Delta}(x)
\right] \cr
&=& \frac{\Sigma^2}{F^2}\Delta(x) + O(m_{\pi}^2) ~,
\eeqn
which is precisely the tree-level result of the $p$-expansion. The constant
again provides the term needed to restore the pion zero-mode
 term in the propagator.

\noi
Again it is convenient to reformulate these results in terms of projections
onto zero momentum. The relevant definitions have already been given
above, and in Figure ~\ref{fig:ppns} we show ${\cal P}^a(t)$ for
 $\nu=0, 1, 2$ and various quark masses within the $\epsilon$ regime.
As expected, the effect of the zero modes are dominant in this regime,
inducing a large dependence on the topological charge.
 The same effect is observed for the scalar correlator in Fig.~\ref{fig:ssns}.
Note that this correlation function is negative for $\nu=1,2$. This feature was first noted by the authors of \cite{Bardeen}  for
larger quark masses in the $p$-expansion regime (ie. not at fixed topology).
As explained above, it is possible to cancel the poles induced by the
topological zero modes by considering the sum of the scalar and pseudoscalar
propagators. This is shown in Figures~\ref{fig:spns}. The only divergence in
the $\mu\rightarrow 0$ limit is then restricted to the $\nu=0$ sector and it
is only logarithmic.

\begin{figure}
\vspace{0.0cm}
\hspace{-0.0cm}
\centering{\epsfig{file=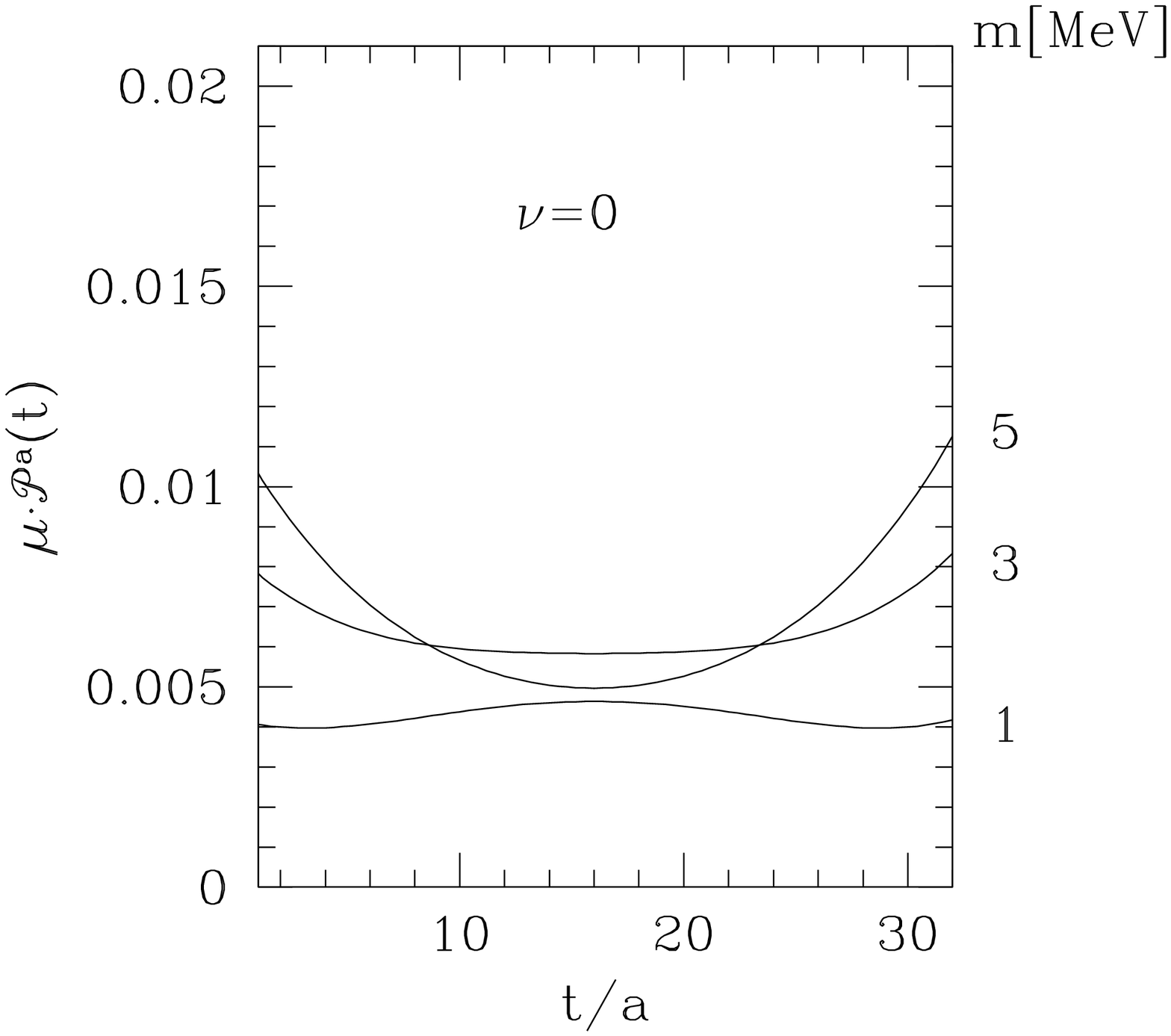,height=9cm,width=12cm}} \\
\epsfig{file=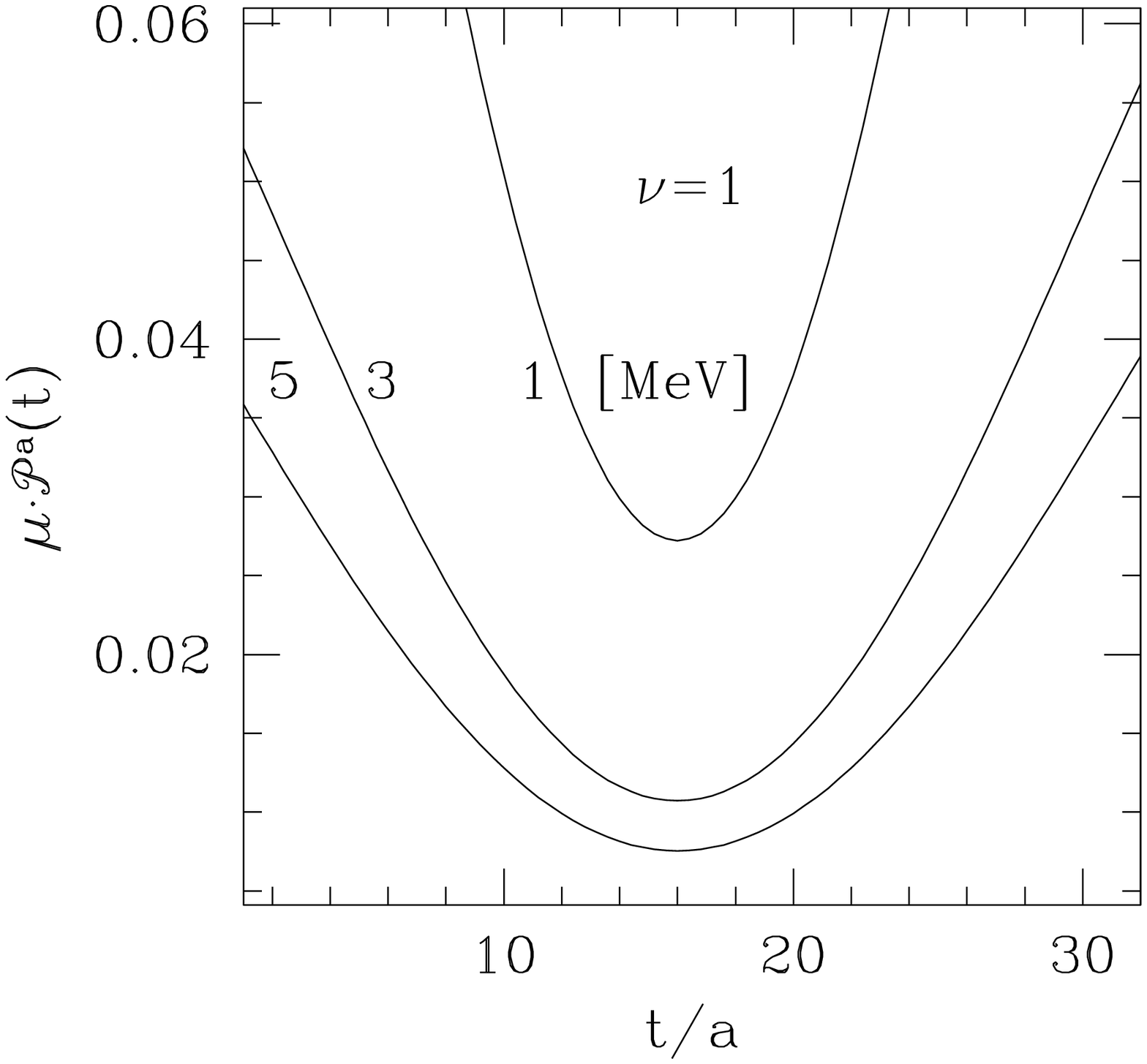,height=9cm,width=8cm}
\epsfig{file=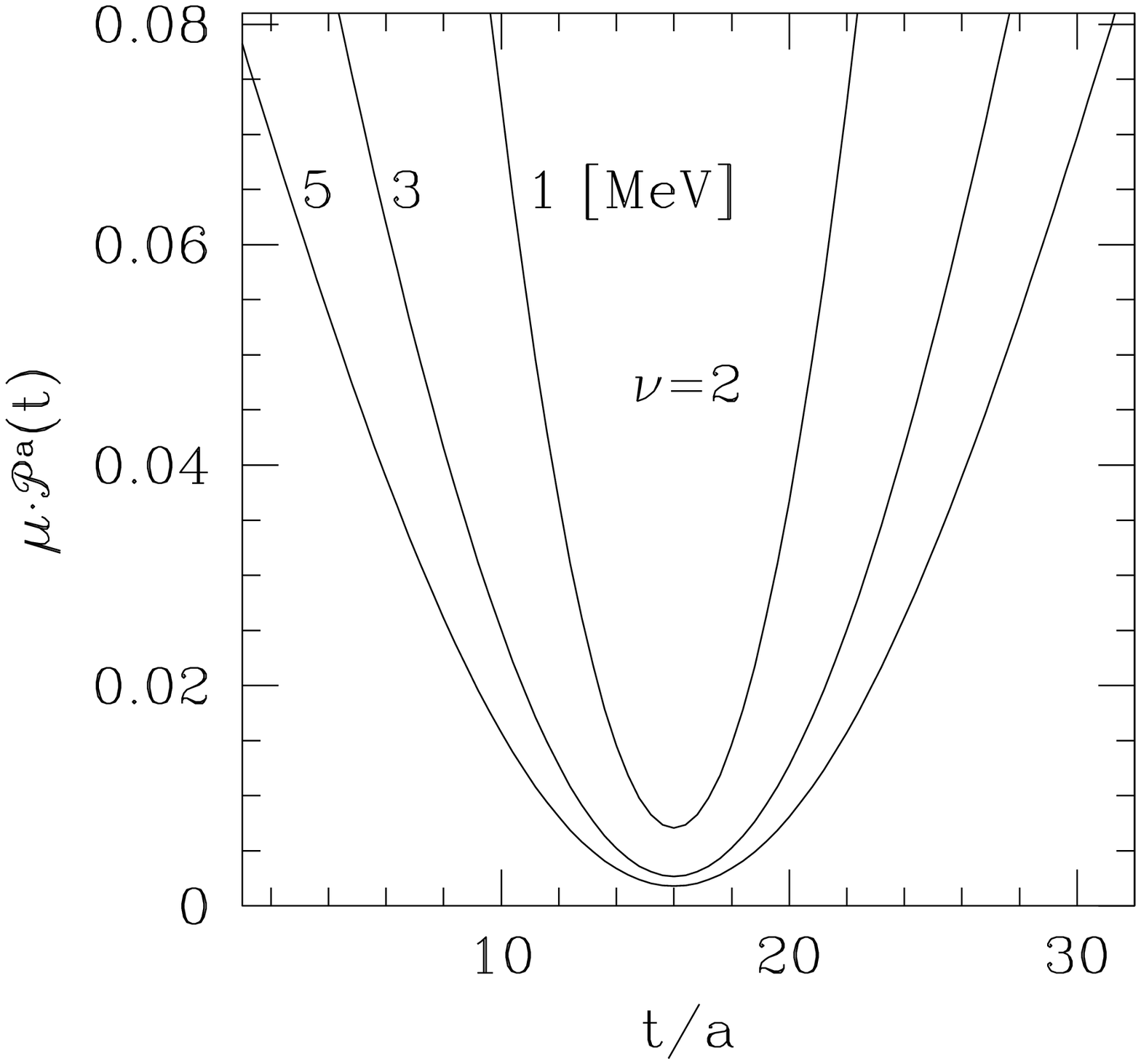,height=9cm,width=8cm}
\caption{ \label{fig:ppns} $\mu {\cal P}^a(t)$ for quark masses $m=1, 3$ and 5 MeV and $\nu=0$ (upper), 1 (down left), 2 (down right). The lattice volume
is $V=L^3 T = 16^3 32$ and $a^{-1} \simeq 2$ GeV.}
\end{figure}

\begin{figure}
\vspace{0.0cm}
\hspace{-0.0cm}
\centering{\epsfig{file=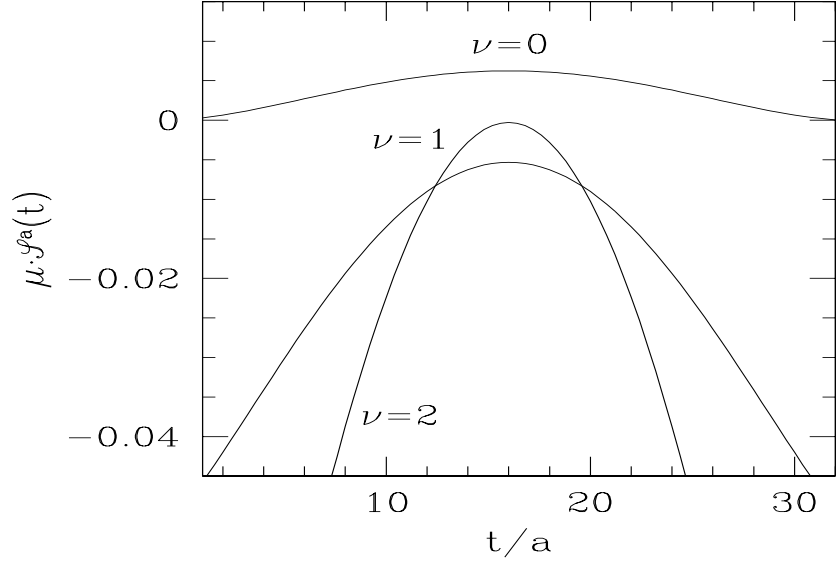,height=9cm,width=12cm}}
\caption{ \label{fig:ssns} $\mu {\cal S}^a(t)$ for a quark mass $m=3$ MeV
and $\nu=0$, 1, 2. The lattice volume is $V=L^3 T = 16^3 32$ and
$a^{-1} \simeq 2$ GeV.}
\end{figure}

\begin{figure}
\vspace{0.0cm}
\hspace{-0.0cm}
\epsfig{file=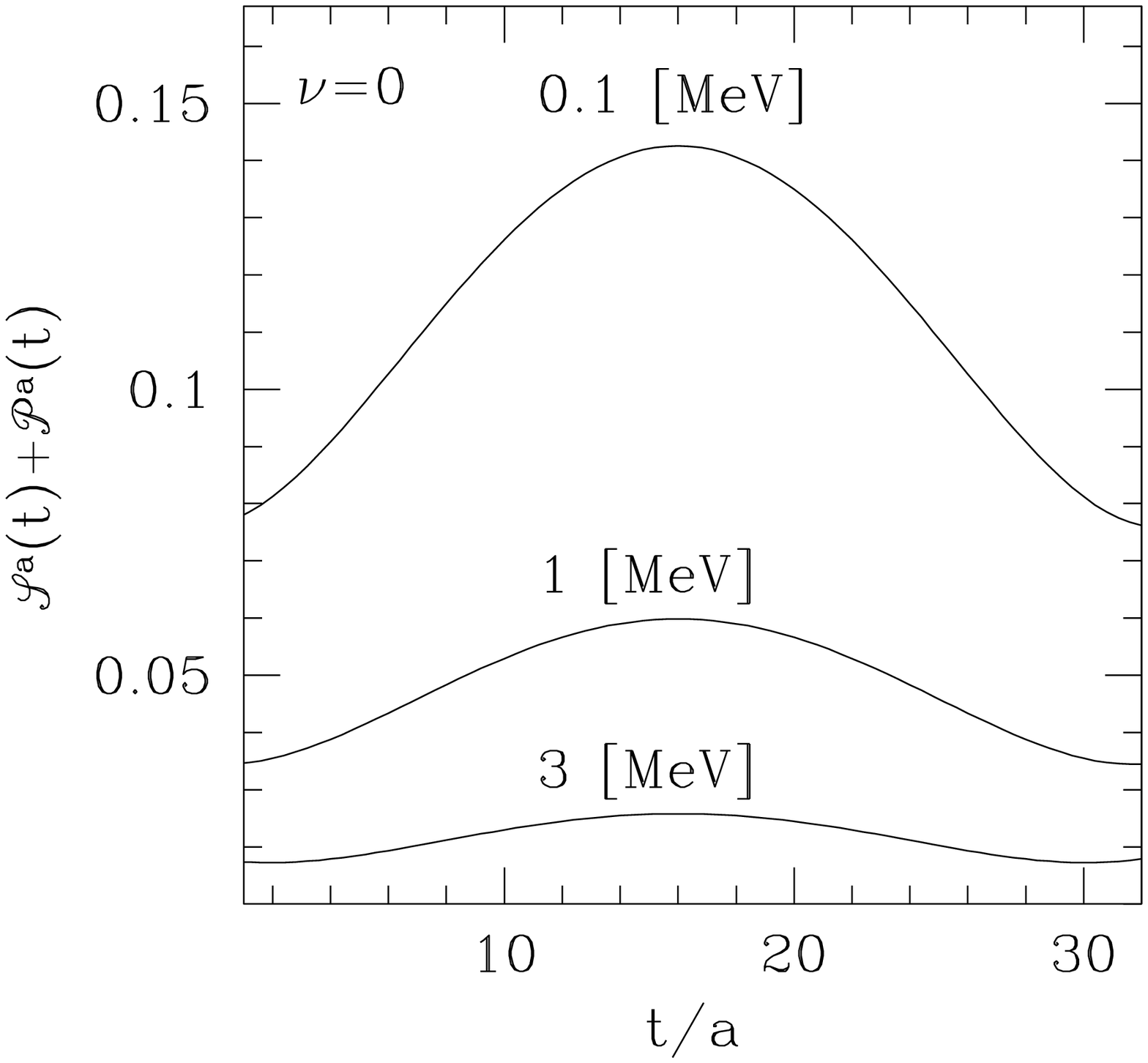,height=9cm,width=8cm}
\epsfig{file=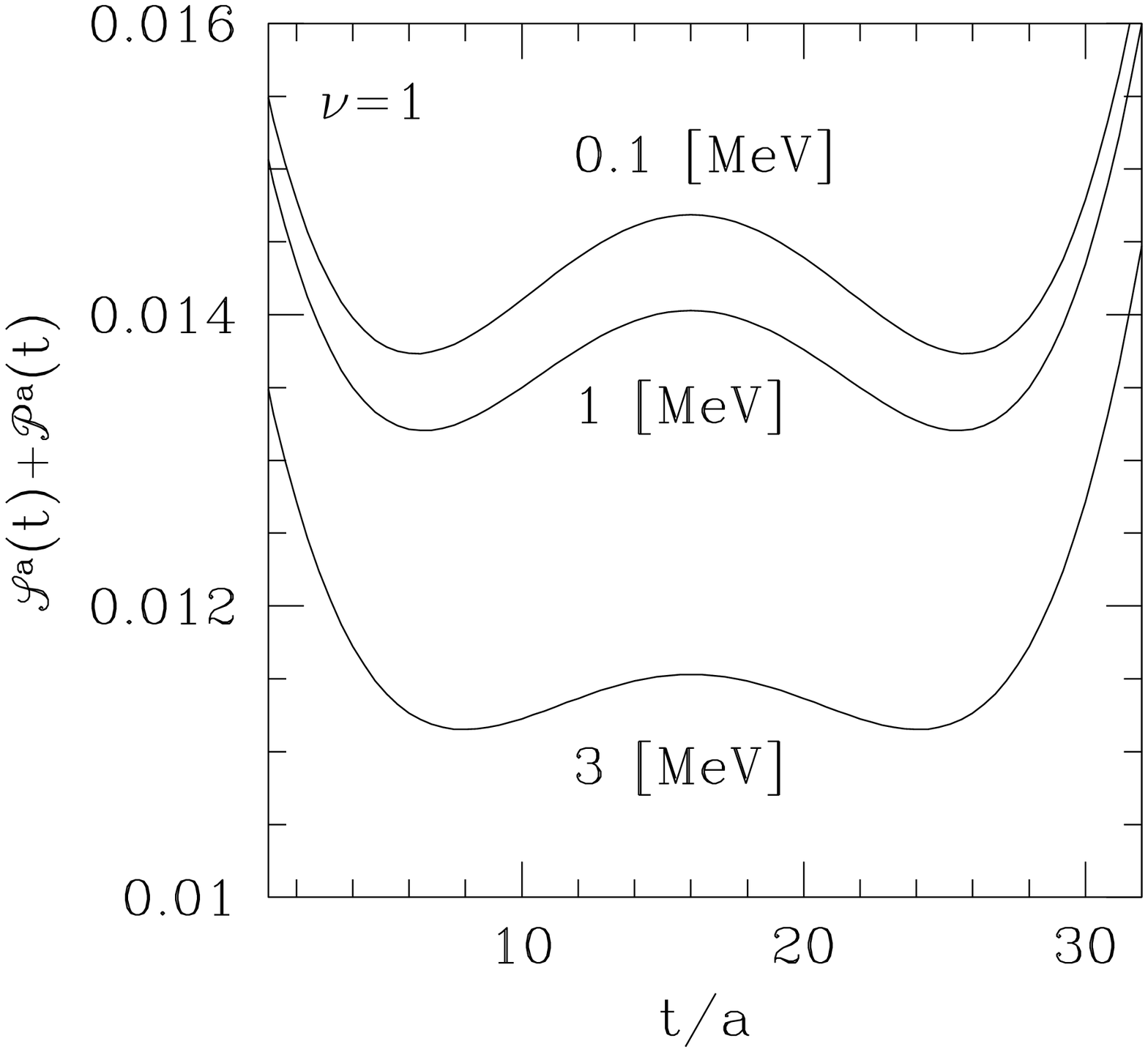,height=9cm,width=8cm}
\caption{ \label{fig:spns} ${\cal P}^a(t) +{\cal S}^a(t)$ for quark masses $m=0.1, 1$ and 3 MeV and $\nu=0$ and 1. The lattice volume
is $V=L^3 T = 16^3 32$ and $a^{-1} \simeq 2$ GeV.}
\end{figure}

\noi
As far as we are aware, there are at present no lattice data for
any of these correlation functions in quenched QCD
at fixed topology and in the appropriate quark mass range.
Very recently, there has been
a study of pseudoscalar correlation functions in the
$N_f=2$ Schwinger model at fixed topology \cite{Durr},
and distinct signatures of
topology were actually found there. But, unfortunately, neither the present
calculation for quenched QCD nor the one presented in section \ref{fullQCD}
for full QCD at fixed topology can directly be compared with these
two-dimensional results. It has also recently been suggested to study
QCD at fixed topology, but at volumes so large that correlations are
still exponential in the masses \cite{BCNW}.

\subsection{{\sc Ward Identities at Fixed Topology}}

Ward identities associated with
the chiral rotation of the physical (valence in the replica context)
quarks provide useful cross-checks on the above results.
For the singlet chiral transformation, the relation is
\beqn
\langle (\partial_\mu A_\mu^0(x) - 2 m P^0(x) - 2 i N_v \omega(x))
{\cal O}(0)\rangle ~=~ -\langle \delta {\cal O}(0) \rangle \delta(x),
\label{wi}
\eeqn
where $A_\mu^0 =\bar{\psi} I_{N_v}  \gamma_\mu \gamma_5 \psi$.  The operator
\beqn
\omega(x) ~=~ \frac{1}{16 \pi^2} Tr F_{\mu\nu} {\tilde F}_{\mu\nu}
\eeqn
is the topological charge density, ${\cal O}(x)$ is any local operator and
$\delta {\cal O}(x)$ is its variation under an
infinitesimal singlet chiral rotation at point $x$. The identity of eq.~(\ref{wi}) is normally
considered in the full theory, after summing over all topological sectors.
But it is exact, and in fact simpler, also in sectors of fixed topology.
We also remark that although all effects of quark loops disappear in the
quenched theory, the above identity reflects the consequences of a chiral
rotation in part of the theory only (i.e. in the valence sector). It would reduce to a triviality
if we were to rotate all fields simultaneously.

\noi
If we combine the Ward identity for ${\cal O}(x) = P^0(x)$ and
${\cal O}(x) =\omega(x)$,
and integrate over space-time, we can eliminate
$\langle\omega(x)P^0(0)\rangle$ to get at fixed topology
\beqn
\int\!d^4 x~ \langle P^0(x) P^0(0) \rangle ~=~
- \frac{\nu^2}{m^2 V} - \frac{\langle S^0
\rangle}{m V}\ .
\label{ppwi}
\eeqn
where $S^0$ is defined in eq.~(\ref{s0p0}) and we have here used
$N_v=1$. Inserting eq.~(\ref{QP}) and taking
into account eq. (\ref{barvanish}), we get,
to $O(\epsilon^2)$:
\beqn
C_P^0 ~=~ \frac{\Sigma^{1-loop}_\nu(\mu)}{m V} - \frac{\nu^2}{(m V)^2} \ ,
\label{CP0QWI}
\eeqn
which coincides precisely with our result in eq.~(\ref{cpcss}). Actually,
at fixed gauge field topology, $\langle\omega(x)P^0(0)\rangle$ can be computed
directly using eq.~(\ref{P0identity}) of Appendix \ref{gl22}:
\beq
\frac{1}{\Sigma}\langle P^0(0) \rangle ~=~
\frac{i\nu}{\mu} ~.
\eeq
Inserting this into the identity of ${\cal O}(x) = P^0(x)$ gives precisely
the same relation as (\ref{CP0QWI}).

\noi
Similarly, the general Ward identity for a flavour non-singlet chiral
rotation is
\beqn
\langle (\partial_\mu A_\mu^a(x) - 2 m P^a(x) ) {\cal O}(0)\rangle
~=~ -\langle\delta^a {\cal O}(0) \rangle \delta(x) ~,
\label{wins}
\eeqn
where $\delta^a {\cal O}(x)$ is the variation of ${\cal O}$ under an 
infitesimal chiral rotation at $x$ in the flavour direction $a$.
Combining the identities thus obtained for ${\cal O}(x) = P^a(x)$ and
${\cal O}(x)=\partial_\mu A_\mu^a(x)$, and subsequently
integrating over space-time we arrive at the relation, to order $\epsilon^2$
(no sum on $a$):
\beqn
\int\! d^4 x~ \langle P^a(x) P^a(0) \rangle ~=~ - \frac{1}{4}
\frac{\langle S^0
\rangle}{m V} ~=~ \frac{1}{2}
\frac{\Sigma^{1-loop}_\nu(\mu)}{m} + ... \ \ ,
\label{ppwi'}
\eeqn
which also agrees with the constant term $C^a_P$ of
eq.~(\ref{cpcsns}), computed to the same order. Note that $S^0$ in
eq.~(\ref{ppwi'}) is the one corresponding to $N_v=2$.

\noi
Although not Ward identities as such, a series of relations that follow from
spectral representations of the involved correlation functions can
also give useful checks on our results. Such relations were derived in
ref. \cite{EHN}, using the language of lattice overlap fermions, but they
can easily be transcribed into continuum language. In the quenched case
a particularly interesting relation gives a sum rule for the flavoured
scalar correlation function. In our normalization, with no sum on $a$,
\beq
\int\! d^4 x~  \langle S^a(x) S^a(0) \rangle ~=~ -\frac{1}{2V}
\frac{\partial}{\partial m}\langle S^0
\rangle ~=~ \frac{1}{2}\frac{\partial}{\partial m}
\Sigma^{1-loop}_{\nu}(\mu) + \ldots \ \ ,
\label{strange}
\eeq
which agrees with our value of the constant $C^a_S$ in eq. (\ref{cpcsns}).
The authors of ref. \cite{EHN} have also derived some general relations
for the $m \to 0$
limit of the finite volume correlation functions we have considered
here. For $N_v=2$ they are in our normalization (again no sum on $a$):
\beqn
\int\!d^4 x~ \left[\langle P^0(x) P^0(0) \rangle -
4\langle S^a(x) S^a(0) \rangle\right] &\sim & \frac{4|\nu|}{m^2V}
- \frac{4\nu^2}{m^2V} \label{P0-Sa} \\
\int\!d^4 x~ \left[\langle P^a(x) P^a(0) \rangle -
\langle S^a(x) S^a(0) \rangle\right] &\sim & \frac{|\nu|}{m^2V}
\label{Pa-Sa}
\eeqn
as $m \to 0$. We have not computed $\langle P^0(x) P^0(0) \rangle$
explicitly for $N_v=2$, but to check eq.~ (\ref{P0-Sa}) we need
only the constant part of this $N_v=2$ correlator. It is fixed by the Ward
identity of eq.~(\ref{wi}) to be
$$
N_v\left[\frac{\Sigma_\nu(\mu)}{m V} - \frac{N_v\nu^2}{(m V)^2}\right]
$$
to lowest order. Plugging this in, and using the relation eq.~(\ref{strange}), which
we have just confirmed, we find that eq.~(\ref{P0-Sa}) is indeed satisfied.
Similarly, eq.~(\ref{Pa-Sa}) is easily seen to be satisfied by
taking the $m \to 0$ limit of our exact finite-volume expressions.

\section{Full QCD at Fixed Topology}
\label{fullQCD}

Eventually lattice computations with light quarks will be pushed beyond the
quenched limit. It is therefore useful to know the corresponding
analytical expressions for finite-volume meson propagators also in
full QCD. The $\epsilon$-expansion of the chiral Lagrangian
is much simpler in this case, as all effects of the flavour singlet
term can be ignored. More precisely, we obtain the relevant chiral
Lagrangian by taking the limit $m_0 \to \infty$ of the replica chiral
Lagrangian. This
enforces $\Tr\xi(x)=0$, and the singlet field decouples.
The relevant expansion was carried through to order
$\epsilon^4$ in ref. \cite{H}, probably beyond the
realistic accuracy of lattice simulations in the near future.
     It is of interest to find also the
corresponding analytical expressions in sectors of fixed gauge field
topology. Such expressions cannot immediately be inferred from ref.
\cite{H}, as all results are listed in a manner valid only for flavour
group $SU(N_f)$\footnote{In particular, the author of ref. \cite{H} has
implicitly used the identity $\langle\Tr(U)\rangle =
\langle\Tr(U^{\dagger})\rangle$, which holds only after summing over
topological charges. Also, other identities specific to the group $SU(N_f)$
have been employed prior to the listing of results in ref. \cite{H}.}.
We first list the correlation functions
up to order $\epsilon^2$ in the full theory with summation
over topological charges, which agree
with the results of ref. \cite{H} to this order\footnote{Our notation is the same as in the
previous sections.}:
\beqn
\langle S^0(x)S^0(0) \rangle &=& C_S^0 -\frac{\Sigma^2}{2F^2}\left[
\left\langle\Tr[(U-U^{\dagger})^2]\right\rangle
- \frac{1}{N_{f}}\left\langle(\Tr(U-U^{\dagger}))^2\right\rangle \right]
\bar{\Delta}(x)         \cr
\langle P^0(x)P^0(0) \rangle &=& C_P^0 + \frac{\Sigma^2}{2F^2}\left[
\left\langle\Tr[(U+U^{\dagger})^2]\right\rangle
- \frac{1}{N_{f}}\left\langle(\Tr(U+U^{\dagger}))^2\right\rangle\right]
\bar{\Delta}(x)       \cr
\langle S^a(x)S^b(0) \rangle &=& \frac{\delta^{ab}}{N_f^2-1}\left\{C_S
-\frac{\Sigma^2}{4F^2}
\left[\left(\frac{N_f^2+1}{N_{f}^2}\right)\left\langle(\Tr(U))^2
+(\Tr(U^{\dagger}))^2\right\rangle\right.\right.\cr
&& \left.\left.- \frac{2}{N_{f}}\left\langle\Tr(U^2) +
\Tr(U^{\dagger 2})\right\rangle
-2N_f^2+4 -\frac{2}{N_{f}^2}\left\langle\Tr(U)
\Tr(U^{\dagger})\right\rangle\right]\bar{\Delta}(x)\right\}         \cr
\langle P^a(x)P^b(0) \rangle &=& \frac{\delta^{ab}}{N_f^2-1}\left\{C_P
+ \frac{\Sigma^2}{4F^2}
\left[\left(\frac{N_f^2+1}{N_{f}^2}\right)\left\langle(\Tr(U))^2
+(\Tr(U^{\dagger}))^2\right\rangle\right.\right.\cr
&& \left.\left.- \frac{2}{N_{f}}\left\langle\Tr(U^2) +
\Tr(U^{\dagger 2})\right\rangle
+2N_f^2-4 +\frac{2}{N_{f}^2}\left\langle\Tr(U)
\Tr(U^{\dagger})\right\rangle\right]\bar{\Delta}(x)\right\} ~.  \label{full}
\eeqn

\noi
To the order at which we are working, we need the
$O(\epsilon^2)$ contributions
to the constant terms. As in the quenched theory, these are entirely
given by the one-loop correction to the chiral condensate, which here
reads \cite{GL}
\beq
\frac{\Sigma_{eff}(V)}{\Sigma} = 1 + \frac{N_f^2-1}{N_f}\frac{1}{F^2}
\frac{\beta_1(L_i/L)}{L^2} + \ldots
\label{Sigmashiftfull}
\eeq
Here $\beta_1(L_i/L)$ is another of the universal shape
coefficients \cite{N,HL}.
To order $\epsilon^2$ the constant terms are given by
\beqn
C_S^0 &=& \frac{\Sigma_{eff}^2}{4}\left\langle(\Tr(U+U^{\dagger}))^2
\right\rangle
\cr
C_P^0 &=& -\frac{\Sigma_{eff}^2}{4}
\left\langle(\Tr(U-U^{\dagger}))^2\right\rangle
\cr
C_S   &=& \frac{\Sigma_{eff}^2}{8}\left[
\left\langle\Tr[(U+U^{\dagger})^2]\right\rangle
- \frac{1}{N_{f}}\left\langle(\Tr(U+U^{\dagger}))^2\right\rangle\right] \cr
C_P   &=& -\frac{\Sigma_{eff}^2}{8}\left[
\left\langle\Tr[(U-U^{\dagger})^2]\right\rangle
- \frac{1}{N_{f}}\left\langle(\Tr(U-U^{\dagger}))^2\right\rangle\right] ~,
\label{fullconst}
\eeqn
where the expectation values are taken with respect to the one-loop improved
action (simply let $\Sigma \to \Sigma_{eff}$ in the tree level action).
As a check on these results,
we note that upon substituting $U \to 1$ the coefficients of $\Delta(x)$
vanish for the first three correlators. Indeed, there
should be no tree level propagation of non-zero modes in any of
these correlation
functions\footnote{In contrast to the quenched case, where we do have
flavour-singlet propagation at tree level.}. A similar substitution in the
last (flavoured, pseudoscalar)
correlation function yields the coefficient $\Sigma^2/F^2$,
corresponding to tree level propagation of the non-zero-momentum
modes in this channel.

\noi
Next, to project down on sectors of fixed topological charge $\nu$
we again integrate over $\theta$ as in section \ref{regimes},
simultaneously extending the
group integration from $SU(N_f)$ to $U(N_f)$ for the zero-momentum
modes. Thus only the integration over the zero-momentum modes is
affected by this projection, and the fluctuation part is precisely
as in ref. \cite{H}. In particular, the fluctuations $\xi^a(x)$ still belong
to the adjoint representation of $SU(N_f)$.
The above expressions then remain valid
in sectors of fixed topological charge, once expectation values over the
zero-momentum modes are
interpreted as being with respect to the measure of
\beq
{\cal Z}_{\nu} ~=~ \int_{U(N_{f})}\! dU~(\det U)^{\nu}\exp\left[\frac{1}{2}\mu
\Tr(U + U^{\dagger})\right] ~. \label{Znudef}
\eeq
One obvious benefit of going to sectors of fixed topological charge
is that all pertinent expectation values can be evaluated explicitly
for any $N_f$.
In order to do so, we first invoke the identities (see Appendix A for
a derivation)
\beqn
\left\langle\Tr(U^2)\right\rangle &=& N_f - \frac{2(N_f+\nu)}{\mu}
\left\langle\Tr (U) \right\rangle \cr
\left\langle\Tr(U^{\dagger 2})\right\rangle &=&
N_f + \frac{2(\nu-N_f)}{\mu}\left\langle\Tr (U^{\dagger})\right\rangle \cr
\left\langle\Tr(U)\right\rangle &=& \left\langle\Tr(U^{\dagger})
\right\rangle  -\frac{2N_f\nu}{\mu} ~.
\label{identities}
\eeqn

\noi
Let us define
\beq
\frac{\Sigma_{\nu}(\mu)}{\Sigma} ~\equiv~ \frac{1}{N_{f}}
\frac{\partial}{\partial \mu}\ln {\cal Z}_{\nu} ~,
\label{Sigmafulldef}
\eeq
for a theory with $N_f$ quarks of equal mass. The partition function
$Z_{\nu}$ is known explicitly in all generality \cite{Brower,LS}.
Up to an irrelevant overall factor the
integration of eq.~(\ref{Znudef}) gives
\beq
{\cal Z}_{\nu}(\{\mu_i\}) =
\det[\mu_i^{j-1}I_{\nu+j-1}(\mu_i)]/\prod_{i>j}^{N_f}(\mu_i^2-\mu_j^2) ~,
\label{integral}
\eeq
where $I_n(x)$ is the modified Bessel function, and the determinant in
the numerator is over a matrix of size $N_f\times N_f$ (the indices $i$ and $j$ denote the matrix elements). In practice
one may be mostly interested in the equal-mass case, where this
expression simplifies considerably. Up to an irrelevant overall
factor,
\beq
{\cal Z}_{\nu}(\{\mu\}) = \det[I_{\nu+j-i}(\mu)] ~,
\eeq
where again the determinant is taken over a matrix of size $N_f\times N_f$.
Then $\Sigma_{\nu}(\mu)$ is known explicitly in terms of modified
Bessel functions for all $N_f$. Next, using
\beq
\frac{\Sigma_{\nu}(\mu)}{\Sigma} ~=~ \frac{1}{2N_{f}}\left\langle
\Tr(U + U^{\dagger})\right\rangle
\eeq
and the last of the identities (\ref{identities}), we obtain
\beqn
\left\langle\Tr(U)\right\rangle
&=& N_f\left(\frac{\Sigma_{\nu}(\mu)}{\Sigma} - \frac{\nu}{\mu}
\right) \cr
\left\langle\Tr(U^{\dagger})\right\rangle
&=& N_f\left(\frac{\Sigma_{\nu}(\mu)}{\Sigma} +
\frac{\nu}{\mu}\right) ~.
\label{newid}
\eeqn
Upon substituting eq.~(\ref{newid}), the first two identities of eq.~(\ref{identities})
as well as eq.~(\ref{finally!}) in eq.~(\ref{full}), we finally get
\beqn
\langle S^0(x)S^0(0) \rangle &=&  C_S^{0(\nu)} + \frac{2\Sigma^2}{F^2}
(N_f^2-1)\frac{1}{\mu}\frac{\Sigma_{\nu}(\mu)}{\Sigma}\bar{\Delta}(x) \cr
\langle P^0(x)P^0(0) \rangle &=&  C_P^{0(\nu)} + \frac{2\Sigma^2}{F^2}
N_f\left[1 - \frac{N_f}{\mu}\frac{\Sigma_{\nu}(\mu)}{\Sigma} - \frac{1}{N_{f}}
\frac{\Sigma_{\nu}'(\mu)}{\Sigma} -\left(\frac{\Sigma_{\nu}(\mu)}{\Sigma}
\right)^2 + \frac{\nu^2}{\mu^2}\right]\bar{\Delta}(x)  \cr
\langle S^a(x)S^b(0) \rangle &=& \frac{\delta^{ab}}{N_{f}^2-1}\left\{
C_S^{(\nu)} - \frac{\Sigma^2}{4F^2}
\left[2N_f\left(\frac{\Sigma_{\nu}'(\mu)}{\Sigma} + N_f\left(
\frac{\Sigma_{\nu}(\mu)}{\Sigma}\right)^2 - N_f\right)\right.\right. \cr
&& + \left.\left.(2N_f^2-4)\frac{\nu^2}{\mu^2}
 + \left(\frac{6N_f^2-4}{N_{f}}\right)\frac{1}{\mu}
\frac{\Sigma_{\nu}(\mu)}{\Sigma}\right]\bar{\Delta}(x)\right\} \cr
\langle P^a(x)P^b(0) \rangle &=& \frac{\delta^{ab}}{N_{f}^2-1}\left\{
C_P^{(\nu)} + \frac{\Sigma^2}{4F^2}
\left[2(N_f^2+2)\left[\left(\frac{\Sigma_{\nu}(\mu)}{\Sigma}\right)^2
+ \frac{1}{N_{f}}\frac{\Sigma_{\nu}'(\mu)}{\Sigma}\right]\right.\right. \cr
&& + \left.\left. 2N_f^2-8
+ \frac{6N_f}{\mu}\frac{\Sigma_{\nu}(\mu)}{\Sigma}-
\frac{8\nu^2}{\mu^2}\right]\bar{\Delta}(x)\right\} ~. \label{fullnu}
\eeqn
As a check of these expressions
we can take the limit $\mu \to \infty$. For any finite $\nu$, this freezes
the zero-mode integral at $U = 1$, and we indeed find that the coefficients of the $\Delta(x)$ terms  in the first three equations of eq. (\ref{fullnu}) vanish in that limit.
Similarly, the last coefficient approaches the required coefficient
$\Sigma^2/F^2$ for reproducing the tree-level
propagation of non-zero-momentum modes of the flavoured Goldstone bosons.

\noi
The explicit expressions for the constant terms in eq.~(\ref{fullnu}) are
\beqn
C_S^{0(\nu)} &=& \Sigma^2 N_f\left[\frac{\Sigma^{1-loop~'}_{\nu}(\mu)}
{\Sigma} +
N_f\left(\frac{\Sigma^{1-loop}_{\nu}(\mu)}{\Sigma}\right)^2\right]   \cr
C_P^{0(\nu)} &=& \Sigma^2 N_f\left[\frac{1}{\mu}\frac{\Sigma^{1-loop}_{\nu}
(\mu)}{\Sigma} - \frac{\nu^2N_f}{\mu^2}\right] \cr
C_S^{(\nu)}   &=&  \frac{\Sigma^2 N_f}{2}
\left[\left(\frac{\Sigma_{eff}}{\Sigma}\right)^2 -
\frac{1}{N_{f}}\frac{\Sigma^{1-loop~'}_{\nu}(\mu)}{\Sigma}
- \frac{N_f}{\mu}\frac{\Sigma^{1-loop}_{\nu}(\mu)}{\Sigma_{\nu}}
- \left(\frac{\Sigma^{1-loop}_{\nu}(\mu)}{\Sigma_{\nu}}\right)^2 +
\frac{\nu^2}{\mu^2}\right]       \cr
C_P^{(\nu)}   &=& (N_f^2-1)\frac{1}{2\mu}\Sigma\Sigma^{1-loop}_{\nu}(\mu) ~.
\label{fullconstants}
\eeqn


\noi
One notices that these QCD correlation functions at fixed topology need
not be positive-definite. In the quenched theory the negative correlations
had an immediate interpretation in terms of unitarity violation caused
by eliminating the fermion determinant. Here, the origin of unitarity
violation is only the restriction to fixed  gauge field topology,
and this peculiarity should disappear once we
sum over $\nu$. Consider the main culprit, the constant part
$C_P^{0(\nu)}$ of the $\langle P^0(x)P^0(0)\rangle$ correlation function.
For any given $N_f$ we will have
\beq
C_P^{0(\nu)} < 0
\eeq
for sufficiently large $\nu$. But when we sum over topology with the correct
weight this does not occur. Consider the large-$\mu$ limit, where calculations
can be made very explicitly. The distribution of winding numbers is then
Gaussian \cite{LS}, with average \cite{DVV}
\beq
\langle \nu^2 \rangle ~=~ m\Sigma V/N_f ~=~ \mu/N_f ~.
\eeq
In this large-$\mu$ limit we can average over topology explicitly,
keeping the leading order in $1/\mu$:
\beqn
\langle C_P^{0(\nu)}\rangle &=& \Sigma^2 N_f\left\langle
\frac{1}{\mu}\frac{\Sigma_{\nu}(\mu)}
{\Sigma} - \frac{\nu^2N_f}{\mu^2}\right\rangle \cr
&\sim & \Sigma^2 N_f\left[
\frac{1}{\mu} - \frac{\langle\nu^2 \rangle N_f}{\mu^2}\right] \cr
&=& 0 ~.
\eeqn
To this order, negativity is precisely just avoided. This is consistent
with the expression of eq.~(\ref{fullconst}), where, after taking the limit
$U \to 1$, we also find zero.

\noi
As in the quenched theory,
some of the above results have nice explanations in terms of general
Ward identities, which  are formally the same as in the quenched theory, with
the obvious replacement $N_v \to N_f$.
Thus, from eq.~(\ref{wi}), by combining again the identities corresponding to
${\cal O}(x) = P^0(x)$ and ${\cal O}(x) = \omega(x)$ we obtain to
order $\epsilon^2$, after integrating over space-time:
\beq
C_P^{0(\nu)} ~=~ \Sigma^2 N_f\left[\frac{1}{\mu}
\frac{\Sigma^{1-loop}_{\nu}(\mu)}
{\Sigma} - \frac{\nu^2N_f}{\mu^2}\right] ~,
\label{CP0Ward}
\eeq
which precisely matches what we found by explicit computations in
eq.~(\ref{fullconstants}). Similarly, the identity for a non-singlet chiral
flavour rotation of eq. (\ref{wins}) requires
\beq
C_P^{(\nu)} ~=~ (N_f^2-1)\frac{1}{2\mu}\Sigma\Sigma^{1-loop}_{\nu}(\mu) ~,
\eeq
to ${\cal O}(\epsilon^2)$
which again agrees precisely with the result of the explicit computation
in eq.~(\ref{fullconstants}).

\noi
Finally, we can also verify the Ward identity of eq.~(\ref{wi}) for ${\cal O}(x) =
P^0(x)$ directly. Using the last of the identities of eq. (\ref{identities})
we find, to this order in the $\epsilon$-expansion:
\beq
\frac{1}{\Sigma}\left\langle\bar{\psi}(x)i\gamma_5\psi(x)\right\rangle =
-\frac{i}{2}\left\langle \Tr(U-U^{\dagger})\right\rangle
= \frac{i\nu N_f}{\mu} ~.
\eeq
Then the Ward identity eq.~(\ref{wi}), after integrating over space-time, is seen
to be identical to eq.~(\ref{CP0Ward}), and hence also verified by our
explicit calculations.

\noi
There is no simple analogy of the quenched relation eq.~(\ref{strange}),
but the other relations of ref. \cite{EHN} for $N_f=2$ written in our
normalization in eqs. (\ref{P0-Sa}) and (\ref{Pa-Sa}) should be satisfied. One first has to compute the
chiral condensate in the $N_f=2$ theory from eq. (\ref{Sigmafulldef}),
and also find its first derivative.
Next, plugging in $N_f=2$ in eq. (\ref{fullconstants}),
and taking the $\mu\to 0$ limit, we find complete
agreement. The cancellations required for this to happen are rather
non-trivial, and hence constitute one more independent check on the
calculation.

\section{Conclusions}
\label{conclude}

There are two very different regimes in which one can take the chiral
limit of finite-volume QCD. The one that is normally considered is the
one in which the volume is required to always exceed by far the Compton
wavelength of all excitations, including the pseudo-Goldstone bosons
that eventually become ultra-light. In this regime, there are still
finite-volume effects, but they are exponentially small. This regime
can be studied by lattice gauge theory simulations, but as the quark
masses are decreased, it becomes increasingly difficult to provide the
enormous volumes that are needed. This is where the usual
$p$-expansion of chiral perturbation theory is relevant. For instance,
if one wishes to measure the pion mass, the straightforward procedure
is to ensure {\em a posteriori} that the linear extent of the
lattice $L$ by far exceeds the inverse pion mass $1/m_{\pi}$, and then
fit the Monte Carlo data to the usual zero-momentum projection of
the pseudoscalar correlation function of eq.~(\ref{finiteVcor}).
Here we are advocating another procedure, which is much less costly
in terms of computer resources close to the chiral limit. Instead of
insisting on the large-volume
condition $L \gg 1/m_{\pi}$, one can go to any lattice size, and
even go to the opposite regime $L \ll 1/m_{\pi}$.
In this regime another chiral perturbation theory, that of the
$\epsilon$-expansion, is relevant. If one measures the same pseudoscalar
correlation function in this regime, there is a modified formula
that describes the correlation function. It depends on the same parameters
of the chiral Lagrangian as in the infinite-volume case, and therefore
it can just as well be used to determine these parameters. In particular,
to lowest order it depends on the infinite-volume chiral condensate
$\Sigma$, the pion decay constant $F$, and the quark mass $m$. By the
Gell-Mann--Oakes--Renner relation, with corrections if required,
this directly provides the pion mass $m_{\pi}$. In this way the pion
mass can be determined from a correlation function in a volume so small
that not one single pion Compton wavelength can fit inside.

\noi
In this paper we have discussed the new analytical formulae that arise in this
regime of the $\epsilon$-expansion. In particular, we have derived the
scalar and pseudoscalar correlation functions for quenched QCD in sectors
of fixed topological charge $\nu$, and for full QCD with $N_f$ light flavours,
also in sectors of fixed topology. Simultaneously
we have discussed various aspects of the quenched
$\epsilon$-expansion, especially its limitations, and the reason
why it is particularly advantageous to consider the chiral limit of
the quenched theory
in sectors of fixed topology. We hope also to have made it apparent that
the alternative replica formulation of quenched chiral perturbation theory
can be very useful.

\noi
As is well known, quenched QCD is a troubled theory, and
in the end one should not consider our predictions for the quenched theory
as more than a testing-ground for full QCD. In particular, we have
found that the quenched finite-volume logarithms of the
$\epsilon$-expansion do appear also in the scalar and pseudoscalar
correlation functions. Here we have simply assumed
that the quenched theory nevertheless can be defined over a certain
range of scales. Our formulae can then be directly compared with results
of quenched computer simulations.

\noi
The predictions of Hansen \cite{H} for full QCD, and our results here
for full QCD in sectors of fixed topology, stand on a
different footing. In full QCD the chiral $\epsilon$-expansion does not suffer
from any of the pathologies of the quenched analogy. This is of course
as expected, since in this case the theory is well-defined. Also here
the analytical predictions can dramatically simplify certain aspects
of Monte Carlo simulations, since they allow for a determination
of the infinite-volume parameters of low-energy QCD from volumes much
smaller than normally thought to be required. It will be interesting to
see also these predictions of full QCD confronted with lattice gauge theory.

\section{Acknowledgement}

We wish to thank L.~Lellouch, M.~L\"uscher and H.~Wittig for useful
discussions.

\appendix
\setcounter{equation}{0}

\section{Quenched Integrals over Zero-Momentum Modes}
\label{gl22}

In this appendix we give the technical details on how to derive all
the $Gl(2|2)$ group integrals needed for the calculation of the flavoured
correlation functions in section \ref{flavoured}. We have already outlined
the idea in the main text. Here we also show how at this level one
can switch between the supersymmetric and replica formulations.

\noi
First a simple observation, most easily explained in the replica formulation.
Since $\det{\cal M}^{-\nu}{\cal Z}_{\nu}$ is a function of ${\cal M}
{\cal M}^{\dagger}$ only, there exist a series of identities among
zero-momentum mode expectation values. For example, all expectation
values that can be obtained including ${\cal M}$ sources such
that both $\det{\cal M}$ and ${\cal M}
{\cal M}^{\dagger}$ remain invariant, are identical. This gives us the first
useful identity. Let $\chi = \Sigma V({\cal M} + J)$, with ${\cal M}$ diagonal.
Two sources introduced by either\footnote{In the supersymmetric formalism
the sources are added in the fermion--fermion block only. In the replica
formalism they are added in the upper left-hand corner, and the matrix is
then supplemented by $N-N_v=N-2$ additional mass-degenerate entries.}
\beq
J ~=~ \frac{1}{\Sigma V} \left( \begin{array}{cc}
              0  & j \\
              0   & 0
              \end{array}
      \right) ~
\eeq
or
\beq
J ~=~ \frac{1}{\Sigma V} \left( \begin{array}{cc}
              0  & ij \\
              0   & 0
              \end{array}
      \right) ~,
\eeq
share the same $\det\chi$ and $\chi\chi^{\dagger}$. This immediately
leads to one useful identity, which is needed in section \ref{flavoured}:
\beq
\left\langle U_{21}^2 + (U^{-1})_{12}^2\right\rangle ~=~ 0 ~.
\eeq

\noi
Next we note that for a source matrix
\beq
J ~=~ \frac{1}{\Sigma V} \left( \begin{array}{cc}
              j_1 & 0 \\
              0   & j_2
              \end{array}
      \right) ~,
\eeq
we obtain the expectation value of eq.~(\ref{1122}) after differentiating with
respect to
$j_1$ and $j_2$. What was evaluated explicitly in ref. \cite{TV} was
rather the susceptibility
\beqn
\chi_{\nu}(\mu) &\equiv & \left.\frac{\delta^2}{\delta j_{1}\delta j_{2}}
\ln {\cal Z}_{\nu}\right|_{j_{1}=j_{2}=0} \cr
& = & -\mu^2\left(K_{\nu}(\mu)^2-K_{\nu+1}(\mu)K_{\nu-1}(\mu)\right)
\left(I_{\nu}(\mu)^2
- I_{\nu+1}(\mu)I_{\nu-1}(\mu)\right) ~.
\eeqn
We therefore have, after substituting $\Sigma_{\nu}(\mu)$ from eq.
(\ref{zerocon}):
\beqn
\left.\frac{\delta^2}{\delta j_{1}\delta j_{2}}
{\cal Z}_{\nu}\right|_{j_{1}=j_{2}=0} &=& \frac{1}{4}
\left\langle(U_{11} + U^{-1}_{11})(U_{22}+U^{-1}_{22})\right\rangle \cr
&=& \chi_{\nu}(\mu) + \left(\frac{\Sigma_{\nu}(\mu)}{\Sigma}\right)^2 \cr
&=& 1 + \frac{\nu^2}{\mu^2} ~,
\eeqn
where in the last line the remarkable simplification is due to a series
of Bessel function identities.

\noi
To see how we can use this result, and that of eq.~(\ref{1111}), to evaluate
all needed expectation values,
let us, as an example, consider a source matrix of the form
\beq
J ~=~ \frac{1}{\Sigma V} \left( \begin{array}{cc}
              0   & j \\
              0   & 0
              \end{array}
      \right) ~.
\eeq
This provides us with the expectation value
\beq
\left\langle(U_{21}+(U^{-1})_{12})^2\right\rangle ~=~
4\left.\frac{\delta^2}{\delta j^2}
{\cal Z}_{\nu}\right|_{j=0} ~.
\eeq
To evaluate it, we note that the square roots of the eigenvalues of the
corresponding matrix $\chi\chi^{\dagger}$ with $\chi \equiv {\cal M} + J$ are
\beqn
m_1 &=& \frac{1}{\sqrt{2}}\left(2\mu^2+j^2+
(j^4+4m^2j^2)^{1/2}\right)^{1/2} \cr
m_2 &=& \frac{1}{\sqrt{2}}\left(2\mu^2+j^2-
(j^4+4m^2j^2)^{1/2}\right)^{1/2} ~,
\eeqn
and, in the replica formalism, $N-N_v=N-2$ additional degenerate eigenvalues
$\mu$.
By the chain rule, and using that $(\det\chi)^{-\nu}Z_{\nu}$ clearly is
a {\em symmetric} function of the eigenvalues $m_{1,2}$, we get
\beq
\left.\frac{\partial^2}{\partial j^2}{\cal Z}_{\nu}\right|_{j=0} ~\to~
\frac{1}{2\mu}\left.\frac{\partial}{\partial m_{1}}
{\cal Z}_{\nu}\right|_{m_{1}=m_{2}=\mu}
+ \frac{1}{2}\left.\frac{\partial^2}{\partial m_{1}^2}{\cal Z}_{\nu}
\right|_{m_{1}=m_{2}=\mu}
- \frac{1}{2}\left.\frac{\partial^2}{\partial m_{1}\partial m_{2}}
{\cal Z}_{\nu}\right|_{m_{1}=m_{2}=\mu} \ ,
\eeq
where on the right-hand side $Z_{\nu}$ is the partition function of
diagonal sources
\beq
{\cal M}+J ~=~ \frac{1}{\Sigma V} \left( \begin{array}{cc}
              m_1   & 0 \\
              0      & m_2
              \end{array}
      \right) ~.
\eeq
Performing the required differentiations, and substituting eq.~(\ref{1111})
and eq.~(\ref{Sigma'}), we find the compact expression
\beq
\left\langle(U_{21}+(U^{-1})_{12})^2\right\rangle = 2\left[\frac{1}{\mu}
\frac{\Sigma_{\nu}(\mu)}{\Sigma} + \frac{\Sigma_{\nu}'(\mu)}{\Sigma}\right] ~.
\eeq

\noi
Next, consider a purely imaginary source matrix such as
\beq
J ~=~ \frac{1}{\Sigma V} \left( \begin{array}{cc}
              ij_1   & 0 \\
              0      & ij_2
              \end{array}
      \right) ~.
\eeq
With $\chi = \Sigma V({\cal M} +J)$, the eigenvalues of $\chi\chi^{\dagger}$
are either
$\mu^2+j_1^2, \mu^2+j_2^2$ (in the supersymmetric formulation), or
$\mu^2+j_1^2, \mu^2+j_2^2, \{\mu^2\}$ (in the replica method).
In the latter formulation there are
$N-N_v$ degenerate eigenvalues in the last set. In both cases we have
\beq
\det(\chi) ~=~
\left(1+\frac{ij_{1}}{\mu}\right)\left(1+\frac{ij_{2}}{\mu}\right)
\eeq
in the quenched limit. Now consider
\beq
\frac{\partial^2}{\partial j_{1} \partial j_{2}}\left[(\det\chi)^{-\nu}
{\cal Z}_{\nu}\right]_{j_{1}=j_{2}=0} ~=~
-\frac{\nu^2}{\mu^2} - \frac{2i\nu}{\mu}\left.\frac{\partial}{\partial j_{1}}
{\cal Z}_{\nu}\right|_{j_{1}=j_{2}=0} + \left.\frac{\partial^2}{\partial j_{1}
\partial j_{2}}{\cal Z}_{\nu}\right|_{j_{1}=j_{2}=0} ~.
\eeq
Here the right-hand side actually vanishes, as follows from switching to
the square roots of the eigenvalues of $\chi\chi^{\dagger}$. We thus have
\beq
\left.\frac{\partial^2}{\partial j_{1}
\partial j_{2}}{\cal Z}_{\nu}\right|_{j_{1}=j_{2}=0} ~=~ \frac{\nu^2}{\mu^2}
+ \frac{2i\nu}{\mu}\left.\frac{\partial}{\partial j_{1}} {\cal Z}_{\nu}
\right|_{j_{1}=j_{2}=0} ~.
\eeq
To evaluate the remaining single derivative, we can proceed analogously, using
\beq
0 ~=~ \frac{\partial}{\partial j_{1}}\left[(\det\chi)^{-\nu}{\cal Z}_{\nu}
\right]_{j_{1}=j_{2}=0} ~=~ -\frac{i\nu}{\mu} +
\left.\frac{\partial}{\partial j_{1}} {\cal Z}_{\nu}
\right|_{j_{1}=j_{2}=0} ~. \label{P0identity}
\eeq
Putting everything together, we finally have
\beqn
\left\langle(U_{11}-(U^{-1})_{11})(U_{22}-(U^{-1})_{22})\right\rangle
&=& -4\left.\frac{\partial^2}{\partial j_{1}\partial j_{2}} {\cal Z}_{\nu}
\right|_{j_{1}=j_{2}=0} \cr
&=& \frac{4\nu^2}{\mu^2} ~.
\eeqn
These, and those given in section \ref{QCF}, are all the expectation
values we need for evaluating the flavoured correlation functions.

\setcounter{equation}{0}

\section{$U(N)$ Group Integral Identities}

In section \ref{fullQCD} we used some identities for group integrals
over $U(N_f)$ of the zero-mode action eq.~(\ref{Znudef}). These identities, and
others needed if one wishes to go beyond leading order, can be viewed
as Schwinger--Dyson equations on that particular group manifold. Here
we outline how to derive such identities.

\noi
To begin, we introduce the notion of left- and right-handed differentiation
on the group. Let $t^a$ be the generators of the algebra $u(N_f)$
in a given representation. Differentiation on group elements of $U(N_f)$
in the same representation can be defined by either
\beq
F(Ue^{i\epsilon_a t^a}) ~=~ F(U) + \epsilon_a\nabla^a_R F(U) + \ldots ~,
\eeq
for right-handed differentiation, or
\beq
F(e^{i\epsilon_a t^a}U) ~=~ F(U) + \epsilon_a\nabla^a_L F(U) + \ldots ~,
\eeq
for left-handed differentiation. The latter is probably the most intuitive
to work with, and we will use it in what follows. For simplicity of notation
we omit the subscript $L$ from now on. One immediately sees that an
explicit representation of $\nabla^a$ is
\beq
\nabla^a ~=~ i(t^aU)_{ij}\frac{\partial}{\partial U_{ij}} ~.
\label{nabladef}
\eeq
By hermitian conjugation, or by invoking unitarity, one finds the way
$\nabla^a$ acts on hermitian conjugates. For example,
\beq
\nabla^aU^{\dagger} ~=~ -iU^{\dagger}t^a ~.
\eeq
The derivative in eq.~(\ref{nabladef}) satisfies the Leibniz rule
\beq
\nabla^a(FG) ~=~ (\nabla^a F)G + F(\nabla^a G) ~,
\eeq
and the Lie algebra
\beq
[\nabla^a,\nabla^b] ~=~ f^{abc}\nabla^c ~,
\eeq
where $f^{abc}$ are the structure constants.

\noi
Left-invariance of the Haar measure on U$(N_f$) implies that
\beq
\int\! dU~ \nabla^a F(U) ~=~ 0 ~,
\label{total}
\eeq
which, in conjunction with the Leibniz rule, leads to a simple rule
for partial integration.

\noi
Let us now focus on the zero-mode theory of (\ref{Znudef}). Schwinger--Dyson
equations of this theory are obtained by applying the rule (\ref{total})
through an insertion of a set of functions of $U$ in the group integral.
As an example, consider
\beq
0 ~=~ \int\! dU~ \Tr[t^a\nabla^a(F(U)(\det U)^{\nu}e^{S(U)})] ~,
\label{SDexample}
\eeq
where
\beq
S(U) ~\equiv~ \frac{\mu}{2}\Tr(U + U^{\dagger}) ~.
\eeq
We normalize the generators by $\Tr(t^at^b) = \frac{1}{2}\delta^{ab}$, and
we also need the $U(N_f)$ completeness relation $t^a_{ij}t^a_{kl}
= \frac{1}{2}\delta_{il}\delta_{jk}$ in that normalization.
Choosing $F(U)=U$ in (\ref{SDexample}) leads to the identity
\beq
\left\langle\Tr(U^2)\right\rangle ~=~ N_f - \frac{2(N_f+\nu)}{\mu}
\left\langle
\Tr (U) \right\rangle ~, \label{U2}
\eeq
which was used repeatedly in section \ref{fullQCD}. Because the integration
measure in (\ref{SDexample}) is not real when $\nu \neq 0$,
the analogous identity for the hermitian conjugate does not follow
trivially from this. Rather, from choosing $F(U)=U^{\dagger}$ in
(\ref{SDexample}) one finds
\beq
\left\langle\Tr((U^{\dagger})^2)\right\rangle ~=~
N_f + \frac{2(\nu-N_f)}{\mu}\left\langle\Tr (U^{\dagger})\right\rangle ~.
\eeq
As a consistency check we note that it can also be obtained from eq. (\ref{U2})
by taking the hermitian conjugate while simultaneously letting
$\nu \to -\nu$. Next, by considering an identity such as
\beq
0 ~=~ \int\! dU~ \nabla^0[\Tr(U-U^{\dagger})(\det U)^{\nu}e^{S(U)}] ~,
\label{SDexample1}
\eeq
one finds
\beq
\frac{\mu}{2}\left\langle[\Tr(U-U^{\dagger})]^2\right\rangle
~=~ -\left\langle\Tr(U+U^{\dagger})\right\rangle - \nu N_f
\left\langle\Tr(U-U^{\dagger})\right\rangle ~.
\label{complicated}
\eeq
In order to simplify the last term, one can consider the identity obtained
from
\beq
0 ~=~ \int\! dU~ \nabla^0[(\det U)^{\nu}e^{S(U)}] ~,
\label{SDexample2}
\eeq
which gives
\beq
\left\langle\Tr(U-U^{\dagger})\right\rangle ~=~ -\frac{2N_f\nu}{\mu} ~.
\eeq
Inserting this into eq. (\ref{complicated}) gives
\beq
\frac{\mu}{2}\left\langle[\Tr(U-U^{\dagger})]^2\right\rangle
~=~ -2N_f\frac{\Sigma_{\nu}(\mu)}{\Sigma}+ \frac{2\nu^2N_f^2}{\mu} ~,
\label{Udiff2}
\eeq
with $\Sigma_{\nu}(\mu)$ as defined in eq. (\ref{Sigmafulldef}). We also need
to extract individual terms on the left-hand side. Let ${\cal Z}_{\nu}$ be as
defined in eq. (\ref{Znudef}). Then the relation
\beq
\frac{1}{4}\left\langle[\Tr(U+U^{\dagger})]^2\right\rangle
~=~ \frac{1}{{\cal Z}_{\nu}}\frac{\partial^2}{\partial\mu^2}{\cal Z}_{\nu}
\eeq
gives
\beq
\left\langle[\Tr(U+U^{\dagger})]^2\right\rangle
~=~ 4N_f\left[\frac{\Sigma_{\nu}'(\mu)}{\Sigma} + N_f\left(
\frac{\Sigma_{\nu}(\mu)}{\Sigma}\right)^2\right] ~.
\label{Usum2}
\eeq
Adding and subtracting (\ref{Udiff2}) and (\ref{Usum2}) finally provides
the needed identities:
\beqn
\left\langle\Tr(U)\Tr(U^{\dagger})\right\rangle &=&
N_f\left[\frac{\Sigma_{\nu}'(\mu)}{\Sigma} + N_f\left(
\frac{\Sigma_{\nu}(\mu)}{\Sigma}\right)^2 +\frac{1}{\mu}
\frac{\Sigma_{\nu}(\mu)}{\Sigma} - \frac{\nu^2N_f}{\mu^2}\right] \cr
\left\langle(\Tr(U))^2+(\Tr(U^{\dagger}))^2\right\rangle &=&
2N_f\left[\frac{\Sigma_{\nu}'(\mu)}{\Sigma} + N_f\left(
\frac{\Sigma_{\nu}(\mu)}{\Sigma}\right)^2 + \frac{\nu^2N_f}{\mu^2}
- \frac{1}{\mu}\frac{\Sigma_{\nu}(\mu)}{\Sigma}\right] ~.
\label{finally!}
\eeqn

\end{document}